\documentclass[aps,prd,showpacs,twocolumn,superscriptaddress,nofootinbib,preprintnumbers]{revtex4-1} 

\makeatletter
\def\p@subsection{}
\makeatother

\usepackage{color,graphicx,amsmath,amssymb,bm}
\usepackage[colorlinks=true,citecolor=blue,linkcolor=blue,urlcolor=blue,
            backref=false,pdfborder={0 0 0}]{hyperref}
\usepackage[normalem]{ulem}
\usepackage[utf8]{inputenc} 
\usepackage{mathtools}
\usepackage{float}
\usepackage{multirow}
\usepackage[dvipsnames]{xcolor}
\usepackage{mathrsfs}
\usepackage{diagbox}
\usepackage{booktabs}

\newcommand{\be}{\begin{equation}}
\newcommand{\ee}{\end{equation}}
\newcommand{\beqa}{\begin{eqnarray}}
\newcommand{\eeqa}{\end{eqnarray}}

\newcommand{\di}{\mathrm{d}}

\begin{document}

\title{Enhancing Cosmological Constraints from Foreground-Cleaned CMB Maps Using Large-Scale Structure Surveys}

\author{Shu-Fan Chen} 
\email{sc5848@columbia.edu
 } 
\affiliation{Astrophysics Laboratory, Columbia University, New York, NY 10027, USA}
\affiliation{Department of Physics, Columbia University New York, NY 10027, USA}
\affiliation{Department of Astronomy, Columbia University New York, NY 10027, USA}

\author{J.~Colin Hill}
\affiliation{Astrophysics Laboratory, Columbia University, New York, NY 10027, USA}
\affiliation{Department of Physics, Columbia University New York, NY 10027, USA}

\begin{abstract}
Extragalactic foregrounds contaminate cosmic microwave background (CMB) temperature maps at small angular scales and limit their utility for precision cosmology. The internal linear combination (ILC) is a well-known technique for suppressing these contaminants, but residual foreground power remains a limiting factor. Kusiak {\it et al.}~(2023)~\cite{Kusiak:2023hrz} proposed adding galaxy number-density maps as additional ILC channels, exploiting their correlation with the large-scale structure sourcing these foregrounds to suppress contamination. Here we apply this framework to forecast the gains in CMB-based cosmological parameter constraints from near- and next-generation experiments. Using a halo-model foreground pipeline and a Fisher forecast from joint TT+TE+EE power spectra, we quantify the improvement from galaxy-tracer-assisted ILC cleaning across three configurations: enhanced Simons Observatory (SO) with unWISE or Rubin-like galaxy catalogs, and a futuristic CMB-HD configuration with a hypothetical deep galaxy survey. We find that adding galaxy tracers reduces the residual foreground power in the cleaned temperature map by $\sim$4\%, $\sim$22\%, and $\sim$32\% at $\ell\sim10,000$ for the unWISE, Rubin-like, and futuristic samples, respectively. For the overall variance of the cleaned map at $\ell\sim10,000$, it provides 8\%, 24\%, and 17\% improvements for each combination. The resulting reduction in marginalized parameter error bars is modest for the base six-parameter $\Lambda$CDM model: sub-percent for SO+unWISE, rising to $\sim2\%$ for SO+Rubin-like tracer. Including the effective number of relativistic species $N_{\rm eff}$, we find at most 2.2\% improvements for both SO+Rubin-like and CMB-HD+Futuristic tracer. These results establish the expected gains from combining near-term CMB experiments with current and forthcoming large-scale-structure surveys.
\end{abstract}

\maketitle

\section{Introduction}

The cosmic microwave background (CMB) remains the most powerful probe of cosmology, encoding information about the composition, geometry, and perturbation spectrum of the Universe in its temperature and polarization anisotropies. Decades of increasingly precise measurements --- from COBE~\cite{1990ApJ...354L..37M,1992ApJ...396L...1S} to WMAP~\cite{WMAP:2012nax} to Planck~\cite{Planck:2018vyg} and ground-based experiments including the Atacama Cosmology Telescope (ACT)~\cite{2020JCAP...12..047A,AtacamaCosmologyTelescope:2025blo} and South Pole Telescope(-3G)~\cite{SPT-3G:2022hvq,SPT-3G:2025bzu} --- have established $\Lambda$CDM as the standard cosmological model and placed tight constraints on its parameters. Next-generation instruments, most notably the Simons Observatory (SO)~\cite{SimonsObservatory:2018koc,SimonsObservatory:2025wwn,2026JCAP...04..051A}, promise to extend this program to smaller angular scales, higher sensitivity, and a broader range of science targets, including the effective number of relativistic species $N_{\rm eff}$, the sum of neutrino masses, and primordial gravitational waves.
 
Fully realizing this potential, however, requires confronting the extragalactic foreground emission that dominates CMB temperature maps at small angular scales ($\ell \gtrsim 3000$). The main contributors are the cosmic infrared background (CIB), the thermal Sunyaev-Zel'dovich (tSZ) effect from hot gas in galaxy clusters and groups, the kinematic Sunyaev-Zel'dovich (kSZ) effect, and radio point sources. Each of these traces the large-scale distribution of matter and is therefore correlated not only with themselves but also with one another, complicating both power spectrum estimation and parameter inference. While these foregrounds are also correlated with the CMB lensing convergence, this correlation appears only at higher-order statistics and does not affect the primary CMB power spectrum (although it does affect CMB lensing reconstruction).
 
One well-known, efficient approach to mitigating these contaminants is the internal linear combination (ILC) technique, which forms the minimum-variance linear combination of multi-frequency maps subject to a unit-response constraint for the CMB signal~\cite{Eriksen:2004jg,WMAP:2003cmr,Tegmark:2003ve,Delabrouille:2008qd}\footnote{While a full analysis would ideally model multi-frequency cross-spectra between components in the observed CMB maps~\cite{AtacamaCosmologyTelescope:2025blo}, we instead rely on the ILC-cleaned spectra. As demonstrated in Ref.~\cite{Surrao:2024lgo}, this approach efficiently captures information equivalent to the full multi-frequency cross-spectra analysis for the parameters of interest.}. Variants of the ILC additionally deproject components with known spectral energy distributions (SEDs), such as the tSZ or CIB, at the cost of some additional variance~\cite{Remazeilles:2010hq}. While effective, these deprojection techniques rely on accurate knowledge of the foreground SED, which is well determined for tSZ but uncertain for the CIB, whose modified-blackbody emission varies across the galaxy population and across the sky~\cite{Viero:2012uq,Planck:2013wqd,Mak:2016ykk,Lenz:2019ugy}.
 
An alternative strategy is to exploit the correlation between the foreground sources and the large-scale structure (LSS) that hosts them. Since the CIB and tSZ fields both trace the distribution of matter --- albeit with different redshift weighting and scale dependence --- galaxy surveys provide empirical templates for this contamination that do not require a parametric SED model. Ref.~\cite{Kusiak:2023hrz} showed that galaxy number-density maps can be incorporated directly into the ILC channel vector as additional ``frequency'' channels, with their entries in the response vector set to zero so that the CMB signal-preservation constraint is unaffected.\footnote{The integrated Sachs-Wolfe effect violates this assumption, so this method can only be used at $\ell \gtrsim 100$.} The ILC minimization then automatically assigns these channels nonzero weight whenever they are correlated with the foreground-contaminated frequency maps, effectively using the galaxy field as a data-driven cleaning template. This approach is complementary to deprojection: rather than nulling a specified SED, it suppresses whatever foreground power is correlated with the tracer.
 
In this work, we perform a systematic forecast of the gains that galaxy-tracer-assisted ILC cleaning can deliver for near- and next-generation CMB experiments. We build a halo-model foreground pipeline --- incorporating CIB, tSZ, kSZ, and radio sources, calibrated against ACT and Planck measurements --- and compute harmonic-space ILC weights for joint CMB-frequency plus galaxy-channel data vectors. Cross-correlating the resulting cleaned temperature map with observed $E$-mode polarization yields a TE power spectrum with suppressed foreground contamination from the T-leg, which we use as the basis for a joint TT+TE+EE Fisher-matrix forecast. We consider three configurations spanning current and future capabilities: the full, enhanced Simons Observatory (SO) Large Aperture Telescope survey~\cite{SimonsObservatory:2025wwn} paired with the unWISE galaxy catalog~\cite{Krolewski:2019yrv}, SO paired with a Rubin-like deeper survey, and a futuristic CMB-HD~\cite{Sehgal:2019ewc} setup with a hypothetical $10\times$ deeper tracer catalog as a sample-variance limited observation. For each, we evaluate both the improvement in the cleaned temperature power spectrum and the resulting change in marginalized constraints on the six base $\Lambda$CDM parameters and $N_{\rm eff}$.\footnote{See Ref.~\cite{Goldstein:2026iuu} for a recent 2\% constraint on $N_{\rm eff}$.}
 
The remainder of this paper is organized as follows. Section~\ref{sec:ilc} describes the ILC formalism and its extension to incorporate galaxy tracers. Section~\ref{sec:experimental_setup} presents the experimental setup, foreground models, and galaxy-catalog configurations used in the forecast. Section~\ref{sec:forecast_and_discussions} presents the results, including the improvement in the cleaned temperature spectrum and in cosmological parameter constraints. We conclude in Section~\ref{sec:conclusions}. 

Throughout this paper, we adopt the Planck 2018 fiducial cosmology~\cite{Planck:2018vyg}. We also assume the standard Tinker {\it et al.} (2008) halo mass function~\cite{Tinker:2008ff}, Navarro-Frenk-White (NFW) halo density profiles~\cite{Navarro:1996gj}, as well as the concentration-mass relation in Ref.~\cite{Bhattacharya:2011vr} for our halo-model. All computations are implemented with {\tt class\_sz}~\cite{Bolliet:2023eob,Bolliet:2025oqo}\footnote{Code and documentation can be found in \url{https://github.com/CLASS-SZ}} unless otherwise mentioned.

\section{ILC with External Galaxy Catalogs}
\label{sec:ilc}

\subsection{Standard and Constrained ILC}
The standard ILC~\cite{Eriksen:2004jg} constructs a map of a signal of interest by forming the minimum-variance linear combination of multi-frequency maps, subject to a unit-response constraint for the target signal. Working in harmonic space, an ILC CMB temperature map built from $N$ frequency channels can be written as (with implied summation)
\begin{align}
    \hat{T}_{\ell m}^{\rm ILC} = w_\ell^i \hat{T}_{\ell m}^i\,,
\end{align}
where $T_{\ell m}^{i}$ denotes the harmonic-space temperature map at frequency $i$ and $\hat{\cdot}$ indicates that the field is from observation. The ILC weights $w_{\ell}^i$ are obtained by minimizing the variance of the resulting ILC map,
\begin{align}
    \sigma^2_{\hat{T}_{\ell m}^{\rm ILC}\hat{T}_{\ell m}^{\rm ILC}} = w_\ell^i w_\ell^j \left(\hat{R}_\ell\right)_{ij}
\end{align}
with $\hat{R}_\ell$ being the empirical frequency-frequency covariance matrix of the data, estimated within bins of width $\Delta\ell$,
\begin{align}\label{eq:ilc_cov}
    \left(\hat{R}_\ell\right)_{ij} = \sum_{\ell'=\ell-\Delta\ell/2}^{\ell+\Delta\ell/2}\frac{2\ell'+1}{4\pi}\hat{C}_{\ell'}^{ij}\,.
\end{align}

The additional feature of the standard ILC, relative to the unconstrained variance minimization, is the requirement that the signal of interest be recovered with unit amplitude in the final combination: $w_\ell^i a_i=1$, where the vector $a_i$ encodes the spectral response of the target signal at each frequency and $a_i=1$ for the blackbody CMB. Imposing this constraint together with the variance minimization, and solving the resulting optimization problem via Lagrange multipliers, yields a closed-form expression for the weights~\cite{Eriksen:2004jg},
\begin{align}
    w_\ell^i = \frac{\left(\hat{R}_\ell^{-1}\right)_{ij}a_j}{\left(\hat{R}_\ell^{-1}\right)_{km}a_k a_m}\,,
\end{align}
with indices $i$, $j$, $k$, and $m$ running from $1$ to $N$.

The standard ILC constraint ensures signal preservation but places no requirement on how the weights respond to other physical components present in the data. In many applications, however, one or more contaminants with well-characterized spectral signatures must be explicitly removed from the final map. This is accomplished by extending the optimization with a deprojection constraint, which forces the ILC weights to have null response to a component of specified spectral energy distribution (SED) such as CIB and tSZ: $w_\ell^i b_i=0$, where $b_i$ denotes the spectral response of the deprojected component at frequency channel $i$. Unlike the signal-preservation constraint, which fixes the response to unity, this constraint drives the response to exactly zero, removing the corresponding component from the output map at the cost of some additional variance relative to the unconstrained-deprojection case.

\subsection{LSS Tracers as Additional Frequencies}
Following Ref.~\cite{Kusiak:2023hrz}\footnote{Code is publicly available in \url{https://github.com/olakusiak/deCIBing}}, the constrained ILC formalism can be extended to incorporate galaxy catalogs as additional channels alongside the frequency maps. This extension is motivated by the fact that large-scale-structure tracers, such as galaxy number-density maps, trace the same underlying matter distribution as astrophysical foregrounds including the CIB and tSZ effect, and are therefore correlated with the foreground contamination present in the CMB temperature maps. Incorporating these tracers into the channel vector used in the ILC minimization allows this correlation to be exploited for foreground suppression, without requiring an explicit parametric model of the foreground SED.

The $N$-frequency channel vector is extended to include $N_g$ external galaxy maps, $g^1$,..., $g^{N_g}$, such that the full channel vector becomes
\begin{align}
    \boldsymbol{M}_{\ell m} &= \bigl(\underbrace{\hat{T}^1_{\ell m},\ldots,\hat{T}^N_{\ell m}}_{N},\;
          \underbrace{\hat{g}^1_{\ell m},\ldots,\hat{g}^{N_g}_{\ell m}}_{N_g}\bigr)\,.
\end{align}
As the galaxy maps carry no CMB signal, the corresponding entries of the response vector are set to zero,
\begin{align}
    \boldsymbol{a} = \bigl(\underbrace{1,\ldots,1}_{N},\;
                    \underbrace{0,\ldots,0}_{N_g}\bigr)\,,
\end{align}
so the signal-preservation constraint (unit response to the CMB) is unaffected, while the galaxy channels remain otherwise free.\footnote{The ISW effect is correlated with LSS, which would complicate this procedure --- thus, this method only applies at $\ell \gtrsim 100$.}

This extension enlarges the covariance matrix entering the weight calculation to an $(N+N_g)\times(N+N_g)$ block matrix,
\begin{align}
  \hat{R}_\ell =
  \begin{pmatrix}
    \hat{R}_\ell^{TT} & \hat{X}_\ell^{Tg} \\[4pt]
    \left(\hat{X}_\ell^{Tg}\right)^T & \hat{G}_\ell
  \end{pmatrix}\,,
\end{align}
where $\hat{R}_\ell^{TT}$ is the usual frequency-frequency covariance,
$\hat{X}_\ell^{Tg}$ contains the cross-spectra between the frequency maps and the galaxy tracers, and $\hat{G}_\ell$ is the galaxy auto- and cross-covariance (including shot noise). The minimum-variance weights take
the same closed form as before,
\begin{align}
  w_\ell^\alpha
  = \frac{\left(\hat{R}_\ell^{-1}\right)^{\alpha\beta} a_\beta}
         {\left(\hat{R}_\ell^{-1}\right)^{\gamma\delta} a_\gamma a_\delta}\,,
\end{align}
but now with $\alpha,\beta,\gamma,\delta$ running over the full set of
$N+N_g$ channels. 

Because the galaxy maps are correlated with the foreground-contaminated frequency maps while carrying no CMB signal of their own, the ILC variance minimization assigns them nonzero weight. The procedure effectively infers the spatial template of the foreground component directly from the data, via its correlation with the LSS tracer, rather than relying on an assumed SED. The method is thus similar to a deprojection constraint, with the distinction that the contaminant template is determined empirically rather than specified through a parametric frequency dependence, and accordingly the associated SED-deprojection SNR penalty is reduced or eliminated.\footnote{In fact, one can impose an explicit spatial deprojection constraint requiring that the ILC CMB map has zero correlation with the LSS tracers --- see Ref.~\cite{Kusiak:2023hrz} for exploration of this approach.} This property is particularly advantageous for foregrounds such as the CIB, whose SED is both observationally uncertain and spatially varying.

\subsection{Covariance $\hat{R}_\ell$}
We construct the covariance matrix $\hat{R}_\ell$ analytically, using the halo-model approach. More details will be described in Section~\ref{sec:experimental_setup} for each component. We also set $\Delta\ell=20$ in Eq.~\ref{eq:ilc_cov}.

The frequency-frequency block $\hat{R}_\ell^{TT}$ is given by 
\begin{align}
    \left(\hat{R}^{TT}_\ell\right)_{ij} &= C_\ell^{TT,ij} + N_{\ell}^i\delta_D^{ij} + C_\ell^{{\rm radio},ij} + C_\ell^{{\rm kSZ}} \nonumber\\
    &+ \sum_{X,Y\in\{\rm CIB, tSZ\}}C_\ell^{XY,ij}\,,
\end{align}
where $C_\ell^{TT,ij}$ is the lensed CMB temperature power spectrum, $N_\ell^i$ is the frequency-dependent instrumental noise, $C_\ell^{{\rm radio},ij}$ and $C_\ell^{\rm kSZ}$ are the radio and kSZ power spectra (each treated as uncorrelated with the other foreground components), and the last term is the CIB and tSZ auto- and cross-power spectra between frequencies $i$ and $j$. For the frequency-galaxy block, we have 
\begin{align}
    \left(X_\ell^{Tg}\right)_{ia} = C_\ell^{{\rm CIB},i-g,a} + C_\ell^{{\rm tSZ},i-g,a}\,,
\end{align}
for frequency $i$ and the $a$-th galaxy sample $g$. Since the galaxy maps trace the same large-scale structure that sources the CIB and tSZ signal, these cross-spectra are computed self-consistently within the halo-model formalism. Lastly, the galaxy-galaxy block is 
\begin{align}
    \left(\hat{G}_\ell\right)_{ab} = C_\ell^{gg,ab} + N_\ell^{gg,ab}\,,
\end{align}
which includes the auto- and cross-spectra of the galaxy tracers with shot noise added.



\subsection{Cross-Correlation between $\hat{T}^{\rm ILC}$ and $\hat{E}$}
We can further cross-correlate $\hat{T}^{\rm ILC}$ with the E-mode polarization map, $\hat{E}$, from a chosen frequency channel (or a simple coadd of the frequency channels):
\begin{align}
    \hat{C}_\ell^{\hat{T}^{\rm ILC}\hat{E}} = w_\ell^\alpha \hat{C}_\ell^{M^\alpha\hat{E}}\,.
\end{align}
The expectation value for this cross-spectrum is
\begin{align}
    \bigl\langle \hat{C}^{\hat{T}^{\rm ILC} \hat{E}}_\ell \bigr\rangle
    &= \bigl\langle\sum_{i=1}^{N} w^{i}_\ell\,
    \hat{C}^{\hat{T}^{\rm ILC}_i\hat{E}}_\ell
    + \sum_{a=1}^{N_g} w^{N+a}_\ell
    \hat{C}^{\hat{g}^a\hat{E}}_\ell \bigr\rangle\nonumber\\
    &\approx \bigl\langle\sum_{i=1}^{N} w^{i}_\ell\,
    \hat{C}^{\hat{T}^{\rm ILC}_i\hat{E}}_\ell\bigr\rangle \approx C_\ell^{TE}\,,
\end{align}
at leading order.\footnote{Here we neglect polarized radio source Poisson power, which remains undetected in recent ACT and SPT analyses after source masking~\cite{AtacamaCosmologyTelescope:2025blo,SPT-3G:2025bzu}.} We can thus measure the CMB TE power spectrum with reduced error bars, as the CIB and tSZ contributions to the T variance are reduced via the LSS cleaning. This reduction in the $TE$ error bar is not merely qualitative: it follows directly from the structure of the Gaussian covariance matrix that enters the Fisher forecast, which we now describe.

For the covariance matrix $C^{\rm Fisher}$ used for the Fisher forecast, we consider the Gaussian likelihood and use data from $C_\ell^{\hat{T}^{\rm ILC}\hat{T}^{\rm ILC}}$, $C_\ell^{\hat{E}\hat{E}}$, and $C_\ell^{\hat{T}^{\rm ILC}\hat{E}}$. This takes the form:
\begin{widetext}
\begin{align}
\label{eq.cov}
    C^{\rm Fisher}
    = \frac{1}{f_{\rm sky}(2\ell+1) \Delta \ell}\begin{pmatrix}
        2(C_{\ell}^{\hat{T}^{\rm ILC}\hat{T}^{\rm ILC}})^2 & 2(C_\ell^{\hat{T}^{\rm ILC}\hat{E}})^2 & 2C_\ell^{\hat{T}^{\rm ILC}\hat{T}^{\rm ILC}}C_\ell^{\hat{T}^{\rm ILC}\hat{E}} \\
        2(C_\ell^{\hat{T}^{\rm ILC}\hat{E}})^2 & 2(C_\ell^{\hat{E}\hat{E}})^2 & 2C_\ell^{\hat{E}\hat{E}}C_\ell^{\hat{T}^{\rm ILC}\hat{E}} \\
        2C_\ell^{\hat{T}^{\rm ILC}\hat{T}^{\rm ILC}}C_\ell^{\hat{T}^{\rm ILC}\hat{E}} & 2C_\ell^{\hat{E}\hat{E}}C_\ell^{\hat{T}^{\rm ILC}\hat{E}} & (C_\ell^{\hat{T}^{\rm ILC}\hat{E}})^2+C_\ell^{\hat{T}^{\rm ILC}\hat{T}^{\rm ILC}}C_\ell^{\hat{E}\hat{E}}
    \end{pmatrix}
\end{align}
\end{widetext}
The diagonal blocks are the standard auto-covariances, each scaling as the square of the total power in the relevant field(s). The off-diagonal blocks reflect the correlations among different spectra. For all terms involving $C_\ell^{\hat T^{\rm ILC}\hat T^{\rm ILC}}$ or $C_\ell^{\hat T^{\rm ILC}\hat E}$, the variance is reduced by the LSS cleaning, since the ILC weighting suppresses the variance in $\hat T^{\rm ILC}$. This is why the error bars on $TE$ tighten along with those on $TT$, even though the $E$-mode map itself is unaffected by the cleaning.

\section{Experimental Setup}
\label{sec:experimental_setup}

\subsection{Primary CMB}
\label{sec:primary_cmb}
For the primary CMB signal and instrumental noise, we adopt the upgraded, fully populated enhanced Simons Observatory (SO) Large Aperture Telescope Receiver (LATR) as our main experimental setup \cite{SimonsObservatory:2025wwn}\footnote{We modify the publicly available SO noise model to compute the noise of SO. The code can be found in \url{https://github.com/simonsobs/so_noise_models}.}. We restrict our frequency set to the four channels centered at 93, 145, 225, and 280 GHz, omitting the SO low-frequency (LF) channels at 27 and 39 GHz, which are not included in our pipeline.

Noise power spectra are computed using a two-phase survey model that tracks the transition from the partially populated to the fully populated LATR, as described in Ref.~\cite{SimonsObservatory:2025wwn}: 3 years of observations with 7 optics tubes (OTs; 1 LF + 4 mid-frequency (MF) + 2 ultra-high-frequency (UHF)), followed by 6 years with the fully populated 13-OT receiver (1 LF + 8 MF + 4 UHF), for a total survey duration of 9 years. For each phase we adopt a $20\%$ observing efficiency and an additional $85\%$ ``edge'' factor accounting for non-uniform map depth near the survey boundary, following \cite{SimonsObservatory:2025wwn}. The white-noise map depth for a given frequency channel and phase is
\begin{align}\label{eq:noise_level}
  \Delta_\nu^{(p)} = \frac{\mathrm{NET}_\nu}{\sqrt{N_\mathrm{OT}^{(p)}}} \frac{1}{\sqrt{t^{(p)}}}\sqrt{A}\,,
\end{align}
where $\mathrm{NET}_\nu$ is the per-OT noise-equivalent temperature (NET; goal sensitivity, Table~5 of \cite{SimonsObservatory:2025wwn}), $N_\mathrm{OT}^{(p)}$ is the number of OTs covering that channel in phase $p$, $t^{(p)}$ is the effective integration time for phase $p$, and $A$ is the survey area. The two phases are combined in inverse variance, $1/(\Delta_\nu)^2 = \sum_p 1/(\Delta_\nu^{(p)})^2$, to give a single effective white-noise level per channel. We adopt a sky fraction $f_\mathrm{sky} = 0.4$ for the cosmological analysis throughout, and $f_\mathrm{sky}=0.61$ for the noise level computed from Eq.~\ref{eq:noise_level}~\cite{SimonsObservatory:2025wwn}.

The temperature noise power spectrum for channel $\nu$ is modeled as a white component plus an atmospheric red-noise component, beam-deconvolved:
\begin{align}
  N_\ell^{\nu\nu} = \Delta_\nu^2\,\Omega\left[1 + \left(\frac{\ell_{\rm knee,\nu}}{\ell}\right)^{3.5}\right] e^{\,\ell(\ell+1)\sigma_{b,\nu}^2}\,,
\end{align}
where $\Omega = 4\pi f_\mathrm{sky}$ and $\sigma_{b,\nu} = \theta_{\mathrm{FWHM},\nu}/\sqrt{8\ln 2}$ is the beam standard deviation, with $\theta_{\mathrm{FWHM},\nu}$ the beam full width at half maximum (FWHM). The reference (``knee'') multipoles are $\ell_{\rm knee} = \{2100, 3000, 3800, 3800\}$ at $\{93, 145, 225, 280\}$ GHz, respectively. For the dichroic pairs sharing a focal-plane architecture (93$\times$145 and 225$\times$280 GHz), we additionally include an atmospheric cross-correlation,
\begin{align}
  N_\ell^{\nu\nu'} = r_\mathrm{atm} \sqrt{N_{\ell,\mathrm{atm}}^{\nu\nu}\,N_{\ell,\mathrm{atm}}^{\nu'\nu'}}\; e^{\,\ell(\ell+1)(\sigma_{b,\nu}^2+\sigma_{b,\nu'}^2)/2}\,,
\end{align}
with atmospheric correlation coefficient $r_\mathrm{atm}=0.9$~\cite{SimonsObservatory:2018koc}.

\begin{figure*}
    \centering
    \includegraphics[width=0.95\linewidth]{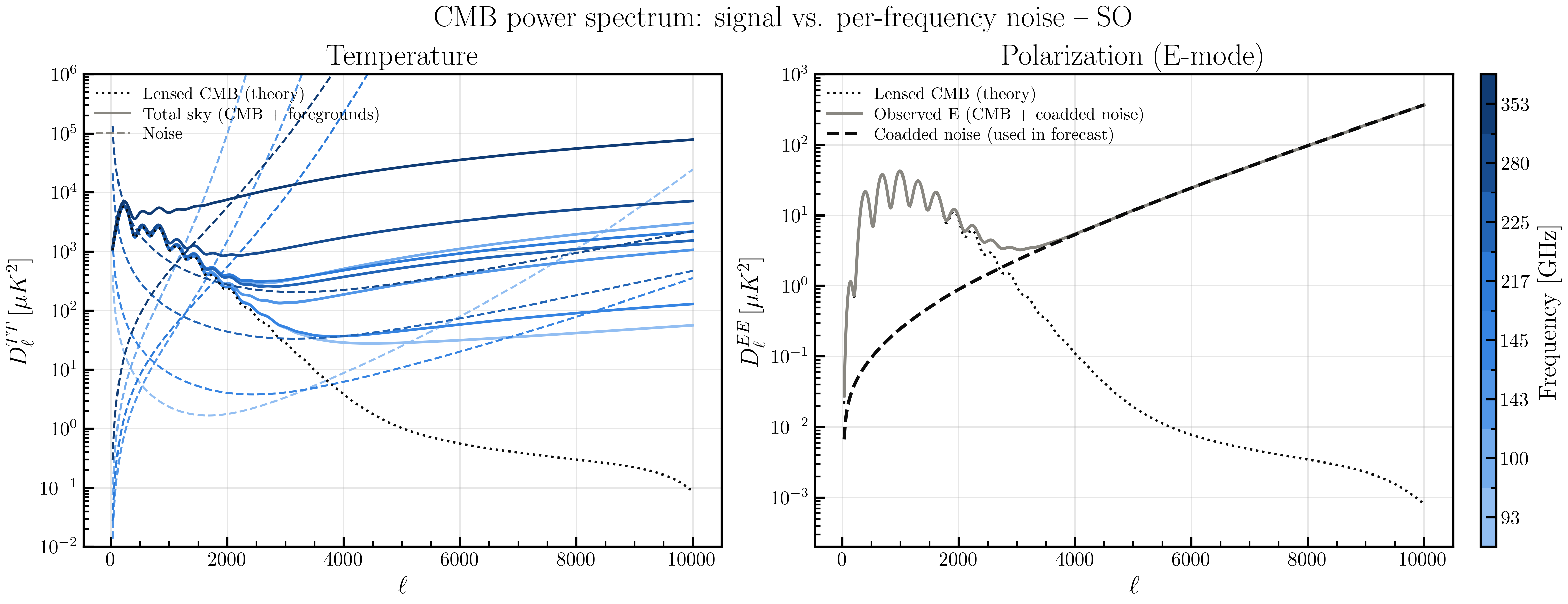}
    \caption{CMB temperature (left) and $E$-mode polarization (right) power spectra for the SO and Planck frequency channels (93--353\,GHz, colorbar). \emph{Left:} solid curves show the total observed sky at each frequency (lensed CMB plus tSZ, kSZ, CIB, and radio foregrounds), and dashed lines show the intrumental noise at each frequency; the black dotted curve is the pure lensed-CMB theory spectrum. 
    \emph{Right:} our pipeline includes no polarized-foreground model, so the black dotted curve is the full $E$-mode signal; the gray solid curve adds the noise actually used in the forecast --- an inverse-variance-weighted coadd of all channels' polarization noise (black dashed).}
    \label{fig:cmb_spectra_SO}
\end{figure*}

The primary CMB TT, TE, and EE power spectra are computed with \texttt{class\_sz} with lensing enabled, up to $\ell_{\rm max} = 10{,}000$, using the Planck 2018 fiducial cosmology~\cite{Planck:2018vyg}\footnote{We increase the precision setting for the computation within {\tt class\_sz} relative to the default setting, including {\tt perturb\_sampling\_stepsize}=0.05, {\tt k\_max\_tau\_over\_l\_max}=15, and {\tt accurate\_lensing}=1, for example. We also follow the precision setup of Ref.~\cite{Kusiak:2023hrz} for the halo-model pipeline so that we can have accurate cross-correlations between different components at small scales.}. In addition to the four SO LAT channels described above, we also include the four Planck channels at 100, 143, 217, and 353 GHz, modeled as uncorrelated white noise following Refs.~\cite{Planck:2018nkj,Planck:2015hzl,Planck:2018vyg,Errard:2015cxa}, to extend the frequency coverage and improve component separation, particularly at lower multipoles where the SO atmospheric noise becomes significant. We set $\ell_\mathrm{max}=2500$ for Planck, and for $\ell\in[30,2500]$ we consider an additional $f_\mathrm{sky}=0.2$ for sky area that does not overlap with SO. We also include large-scale modes $\ell\in[2,29]$ for both T and E, assuming $f_\mathrm{sky}=0.8$~\cite{SimonsObservatory:2018koc}. Figure~\ref{fig:cmb_spectra_SO} shows the signal and the noise power spectra for each frequency.

\subsection{Galaxy Catalog}
\label{sec:galaxy_catalogue}
\begin{figure}
    \centering
    \includegraphics[width=0.9\linewidth]{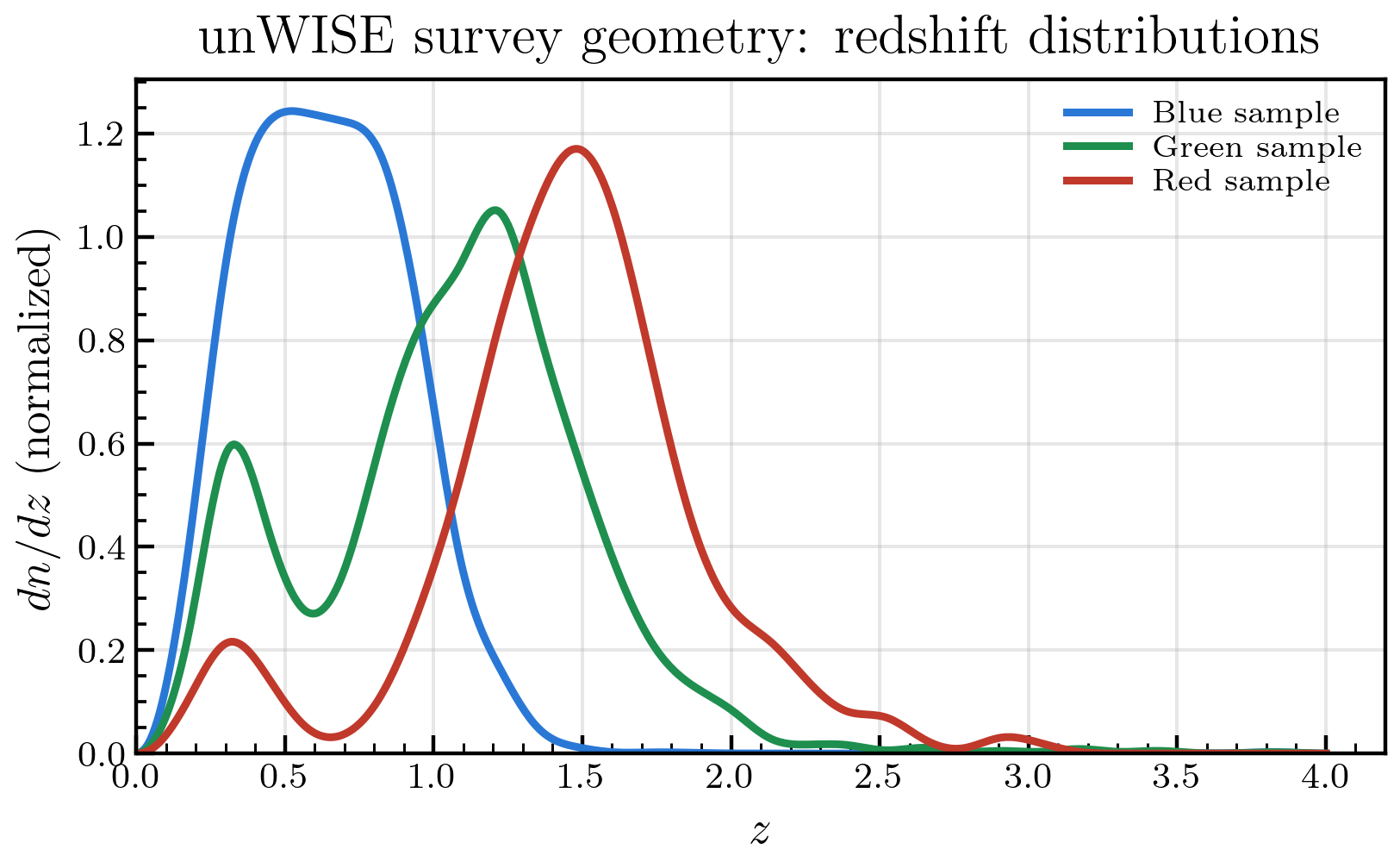}
    \caption{Redshift distributions $\di N/\di z$ of the three color-selected unWISE sub-samples (blue, green, red) used as the galaxy tracer throughout this work, normalized to unit integral. The blue sample peaks at the lowest redshift ($z\sim0.6$) and the red sample at the highest ($z\sim1.5$), with progressively larger high-$z$ tails. This same redshift geometry underlies the Rubin-like and futuristic galaxy-sample configurations of Section~\ref{sec:futuristic_configuration}, which only rescale the shot noise relative to unWISE.}
    \label{fig:unwise_dndz}
\end{figure}

For the galaxy catalog, we consider the unWISE sample~\cite{Krolewski:2019yrv} as our starting point, which has been widely used as a large-scale-structure tracer for CMB lensing reconstruction and cross-correlation studies~\cite[e.g.,][]{Krolewski:2021yqy,Kusiak:2022xkt,ACT:2023oei}. Its sky coverage and depth make it a natural choice for cross-correlating with CMB experiments. The catalog is divided into three color-selected sub-samples --- conventionally labeled blue, green, and red --- constructed from progressively fainter magnitude and color cuts that select galaxies of increasing bias and characteristic redshift. Figure~\ref{fig:unwise_dndz} shows the redshift distribution $\di N/\di z$ of each sub-sample, peaking at $z\sim0.6$, $1.1$, and $1.5$ for the blue, green, and red samples, respectively, with the red sample extending to the highest redshifts and the broadest tail. For parameters of the halo occupation distribution (HOD), we apply the same setup as Ref.~\cite{Kusiak:2023hrz}, which re-fits the unWISE data up to $\ell=4000$ using the modified version of Zheng {\it et al.} (2007)~\cite{Zheng:2007zg,SDSS:2010acc}, together with the DES-Y3 adjustment~\cite{DES:2021olg}. The redshift geometry is fixed across all of the galaxy-sample configurations considered in this work; for simplicity, the Rubin-like and futuristic configurations reuse the same $\di N/\di z$ and differ only in their assumed shot noise levels.\footnote{This is a simplified scenario as HOD parameters should not remain the same with different galaxy number densities.}


We also take into account the galaxy lensing magnification contribution, which modifies the observed galaxy overdensity $\delta_g^{\rm obs}$ as
\begin{align}
    \delta_g^{\rm obs} \rightarrow \delta_g^{\rm obs} + (5s-2)\kappa\,.
\end{align}
Here $s=\di\log_{10}N/\di m$ is the logarithmic slope of the cumulative galaxy number counts at the survey's flux limit, for which we adopt the same values as in Table 1 of Ref.~\cite{Krolewski:2019yrv}, and $\kappa$ is the lensing convergence along the line-of-sight which magnifies background sources and thereby shifts the observed counts relative to the intrinsic ones. Because $\kappa$ is sourced by the same large-scale structure traced by the galaxies, magnification introduces a correlated contribution to not only the galaxy overdensity field itself, but also other components such as the CIB and tSZ fields. 

Table~\ref{tab:unwise_shot_noise} gives the galaxy shot-noise (Poisson) matrix $N^{gg}_{ab}$ used for the auto- and cross-spectra of the three unWISE sub-samples in our forecast. The diagonal terms scale with the inverse number density of each sample, increasing from blue to red as the number density falls. For Rubin-like~\cite{LSST:2008ijt} and the futuristic configuration, we simply divide the shot noise by a factor of $\sim3$ and 10, respectively\footnote{We expect similar results for Euclid~\cite{EUCLID:2011zbd,Euclid:2021icp} and Roman~\cite{Spergel:2013tha,Akeson:2019biv}.}. This is a heavily simplified scenario and we leave accurate modeling of the redshift range and HOD parameters, for example, for future analysis. 

\begin{table}
    \centering
    \begin{tabular}{c|ccc}
        $N^{gg}_{ab}$ & Blue & Green & Red \\
        \hline
        Blue  & $8.70\times10^{-8}$ & $6.22\times10^{-9}$ & $0$ \\
        Green & $6.22\times10^{-9}$ & $1.53\times10^{-7}$ & $4.67\times10^{-8}$ \\
        Red   & $0$                 & $4.67\times10^{-8}$ & $2.88\times10^{-6}$ \\
    \end{tabular}
    \caption{Galaxy shot-noise matrix $N^{gg}_{ab}$ for the auto- and cross-spectra of the three unWISE color sub-samples, as used in the Fisher forecast.}
    \label{tab:unwise_shot_noise}
\end{table}

\subsection{CIB}
\label{sec:cib}
We model the CIB using the halo-model description of the dusty star-forming galaxy (DSFG) population~\cite{Shang:2011mh}, with each galaxy's emission characterized by a modified blackbody SED~\cite{Madhavacheril:2019nfz}. The spectral response of the CIB at frequency $\nu$, relative to a reference frequency $\nu_0=353$ GHz, is
\begin{align}
    b_i^{\rm CIB}(\nu) \propto \left(\frac{\nu}{\nu_0}\right)^{3+\beta}\frac{e^{h\nu_0/k_B T_d}-1}{e^{h\nu/k_B T_d}-1} \times \frac{(\di B_\nu/\di T)\mid_{\nu_0}}{(\di B_\nu/\di T)\mid_\nu}\,,
\end{align}
where $T_d$ and $\beta$ are the effective dust temperature and emissivity index, and the final factor converts from specific intensity to CMB thermodynamic units with $B_\nu$ being the Planck function. In our baseline analysis, we adopt the parametrization of Ref.~\cite{Viero:2012uq}, commonly referred to as the H13 CIB model and also used in the WebSky simulations~\cite{Stein:2020its}, with $T_d=24\,{\rm K}$ and $\beta=1.2$. We do not include the P14 CIB model~\cite{Planck:2013wqd} in our analysis, since it is expected to have limited correlation with the galaxy survey at both large and small scales compared to the H13 model (see Figure 12 of Ref.~\cite{Kusiak:2023hrz} for the comparison).

\begin{figure}
    \centering
    \includegraphics[width=1.0\linewidth]{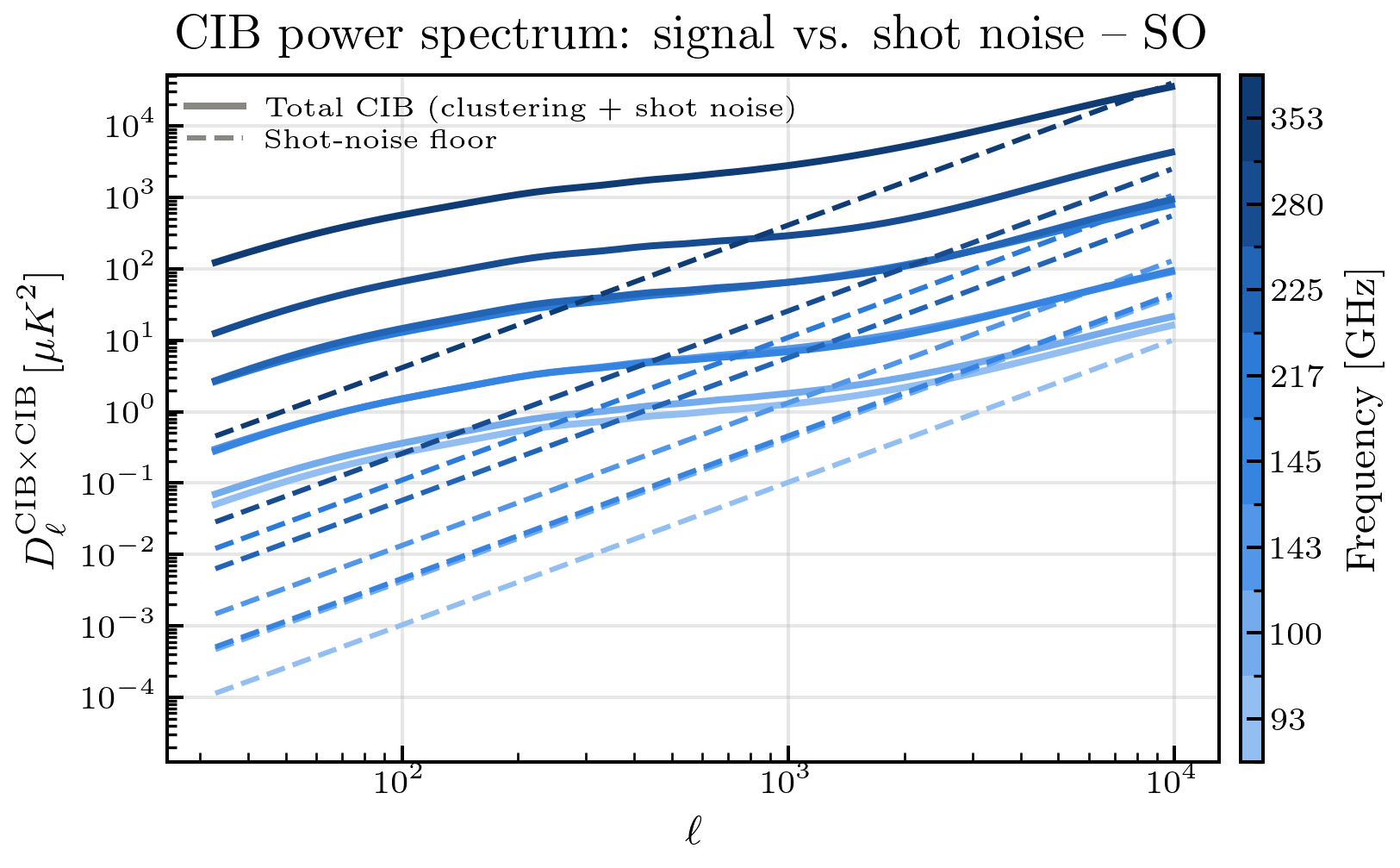}
    \caption{CIB auto-power spectra for the SO and Planck frequency channels (93--353\,GHz). Solid curves show the total CIB spectrum (clustering plus Poisson shot noise) from the \texttt{class\_sz} computation assuming the H13 model; dashed curves show the Poisson shot-noise floor alone. Both components are frequency-dependent, with amplitude increasing toward higher frequency as expected for dusty star-forming galaxies.}
    \label{fig:cib_spectra_SO}
\end{figure}

The diffuse CIB power, decomposed into 1-halo and 2-halo contributions, together with its cross-correlation with galaxy density, follows from the same halo-model ingredients adopted throughout this work, ensuring consistency across all cross-statistics. Because individually detectable galaxies are masked above some flux-density threshold $S_\nu^{\rm cut}$ before the diffuse power is computed, this cut must be specified at each frequency. For the Planck frequencies, we adopt the cuts of the Planck Collaboration~\cite{Planck:2013wqd}. For the SO channels, we adopt the $5\sigma$ point-source detection thresholds reported in Table 3 of Ref.~\cite{SimonsObservatory:2025wwn}: 2.6, 3.35, 7.0, and 12.5 mJy at 93, 145, 225, and 280 GHz, respectively. Galaxies below this flux threshold remain in the map and contribute an additional Poisson shot-noise term to the diffuse power at each frequency pair. We compute this shot noise by integrating the same halo-model luminosity functions used for the diffuse terms, masked at the identical flux cut, and calibrate the result against the measured shot-noise amplitudes from Planck~\cite{Planck:2013wqd} at 100, 143, 217, and 353 GHz. The halo-model prediction is systematically higher than these measurements by a frequency-dependent factor ranging from $\sim 4.5\times$ at 100 GHz to $\sim 3.3$--$3.4\times$ at 217--353 GHz. We divide out this factor, interpolated log-linearly in frequency, which preserves the physical flux-cut and frequency dependence of the halo-model calculation while anchoring its amplitude to the Planck measurements. Figure~\ref{fig:cib_spectra_SO} shows the total CIB power spectrum and the corresponding shot noise for both Planck and SO frequency channels.

\subsection{tSZ}
\label{sec:tsz}
We model the tSZ field using the halo-model pressure profile of Ref.~\cite{Battaglia:2011cq}, specifically the ``AGN feedback'' model evaluated at an overdensity of $\Delta=200$ relative to the critical density. The tSZ spectral response is the standard non-relativistic Compton-$y$ frequency dependence,
\begin{align}
    b_i^{\rm tSZ}(\nu) = T_{\rm CMB}\left(x\coth\frac{x}{2} - 4\right)\,,
\end{align}
where $x\equiv h\nu/k_B T_{\rm CMB}$. Unlike CIB, this frequency dependence has a known and exact functional form, making the tSZ component straightforward to deproject.
 
The Compton-$y$ auto- and cross-power spectra are computed in the halo model using the same mass function, concentration-mass relation, and redshift range adopted throughout this work, integrating the pressure profile over halo mass and redshift. Since the tSZ field traces the same large-scale structure probed by the CIB and by our galaxy catalog, we additionally compute its cross-correlations with both: the tSZ$\times$CIB cross-spectrum, and the tSZ$\times$galaxy cross-spectrum (following the same halo-model prescription, including the corresponding galaxy lensing-magnification contribution, used for the CIB$\times$galaxy cross-correlation). The former enters the frequency-frequency covariance matrix $\hat{R}_\ell$ used in the ILC weight calculation alongside the tSZ and CIB auto-terms; the latter enters the extended covariance block $\hat{X}_\ell^{Tg}$ whenever galaxy maps are incorporated as additional ILC channels, alongside the analogous CIB$\times$galaxy term. Both are computed self-consistently using the same halo-model ingredients as the individual components.

\subsection{Radio Sources}
\label{sec:radio_source}
In addition to the CIB and tSZ contaminants, we include a contribution from Poisson-distributed extragalactic radio point sources. Rather than adopting a single constant-amplitude power law, we model the flux-cut dependence of the radio Poisson power at $\ell_0=3000$ using the double power-law fit of Ref.~\cite{Lagache:2019xto},
\begin{align}
    D_{\ell=\ell_0}^{\rm radio}(\nu,S_{\rm lim})= \frac{2A^{\rm radio}(\nu)}{\left(S_{\rm lim}/S_0(\nu)\right)^{\alpha(\nu)} + \left(S_{\rm lim}/S_0(\nu)\right)^{\beta(\nu)}}\,,
\end{align}
where $S_{\rm lim}$ is the flux-density threshold above which sources are individually masked. The amplitude $A^{\rm radio}$, break flux $S_0$, and power-law indices $\alpha,\beta$ are taken from a fit to an updated Tucci et al.~(2011)~\cite{Tucci:2011nx} source-count model validated against ACT/SPT/Planck number counts, interpolated log-linearly in frequency between tabulated reference values. The full multipole dependence follows the usual Poisson scaling, $D_\ell^{\rm radio} \propto \ell(\ell+1)/[\ell_0(\ell_0+1)]$, and the cross-frequency amplitude between channels $\nu_1$ and $\nu_2$ is taken as the geometric mean $\sqrt{A^{\rm radio}(\nu_1)A^{\rm radio}(\nu_2)}$ of the two single-frequency auto-amplitudes, consistent with a single coherent Poisson-distributed population.
 
We calibrate this model against ACT DR4 measurements~\cite{ACT:2020frw} of the radio Poisson amplitude at $\ell=3000$ and 150 GHz, available at two flux cuts differing by nearly a factor of 7 (15 mJy: $A_{s,d}=3.74\pm0.24\ \mu\mathrm{K}^2$; 100 mJy: $A_{s,w}=22.56\pm0.33\ \mu\mathrm{K}^2$). The raw double power-law-model prediction underpredicts both anchors by a consistent $\sim15$--$20\%$, indicating a flat normalization offset rather than a flux-cut- or frequency-dependent shape error. We therefore apply a single multiplicative correction factor following the same calibration procedure used for the halo-model CIB shot noise (see Section~\ref{sec:cib}).

\subsection{kSZ}
\label{sec:ksz}
For the kSZ signal, we adopt a fixed simulation-based power spectrum template combining the late-time kSZ contribution of Ref.~\cite{Battaglia:2010tm}, sourced by the diffuse post-reionization ionized IGM, with the patchy/reionization-era kSZ contribution of Ref.~\cite{Battaglia:2012im}, sourced by inhomogeneous reionization. This is the same kSZ template family adopted in the ACT DR6 analysis~\cite{AtacamaCosmologyTelescope:2025blo}. Since this is a fixed, cosmology-independent simulation prediction rather than a direct measurement, we rescale the combined template by a single flat multiplicative factor to match the kSZ amplitude constrained by ACT DR6, $a_{\rm kSZ} = 2.0\pm0.9\ \mu\mathrm{K}^2$ at $\ell=3000$ (see Table 7 of Ref.~\cite{AtacamaCosmologyTelescope:2025blo}, or Ref.~\cite{2025JCAP...10..082B}). This preserves the shape of the simulated template while matching its overall amplitude to observations, following the same calibration approach used for the radio source model.

Since the kSZ signal carries no frequency dependence in CMB blackbody temperature units, the same rescaled template is applied identically to every frequency-channel pair entering the covariance matrix $\hat{R}_\ell$. As with the radio source population, we do not model any cross-correlation of the kSZ signal with other sky components (CMB, tSZ, CIB, radio) or with the galaxy catalog (in fact, these cross-correlations vanish for the kSZ field due to the line-of-sight velocity dependence of the kSZ signal), and we do not vary the cosmology of the template.

\subsection{Futuristic Configuration}
\label{sec:futuristic_configuration}
To probe the ultimate reach of this technique, we also consider a futuristic configuration that pairs the CMB-HD instrument noise model~\cite{Sehgal:2019ewc}\footnote{We interpolate the noise level to the frequencies of SO.} with a hypothetical galaxy survey whose number-density shot noise is $10\times$ lower than that of the unWISE catalog. The tracer's shot noise sets a floor on how effectively it can remove extragalactic foreground contamination from the CMB maps. Pushing both the instrumental noise and the tracer shot noise toward their best plausible future values therefore bounds the maximum improvement that galaxy-tracer-assisted cleaning could ultimately deliver, as opposed to the more modest gains expected from any near-term survey pairing. We compare this idealized limit against the SO+unWISE and SO+Rubin-like configurations to gauge how much room remains between current- and next-generation performance.
 
\begin{figure*}
    \centering
    \includegraphics[width=0.9\linewidth]{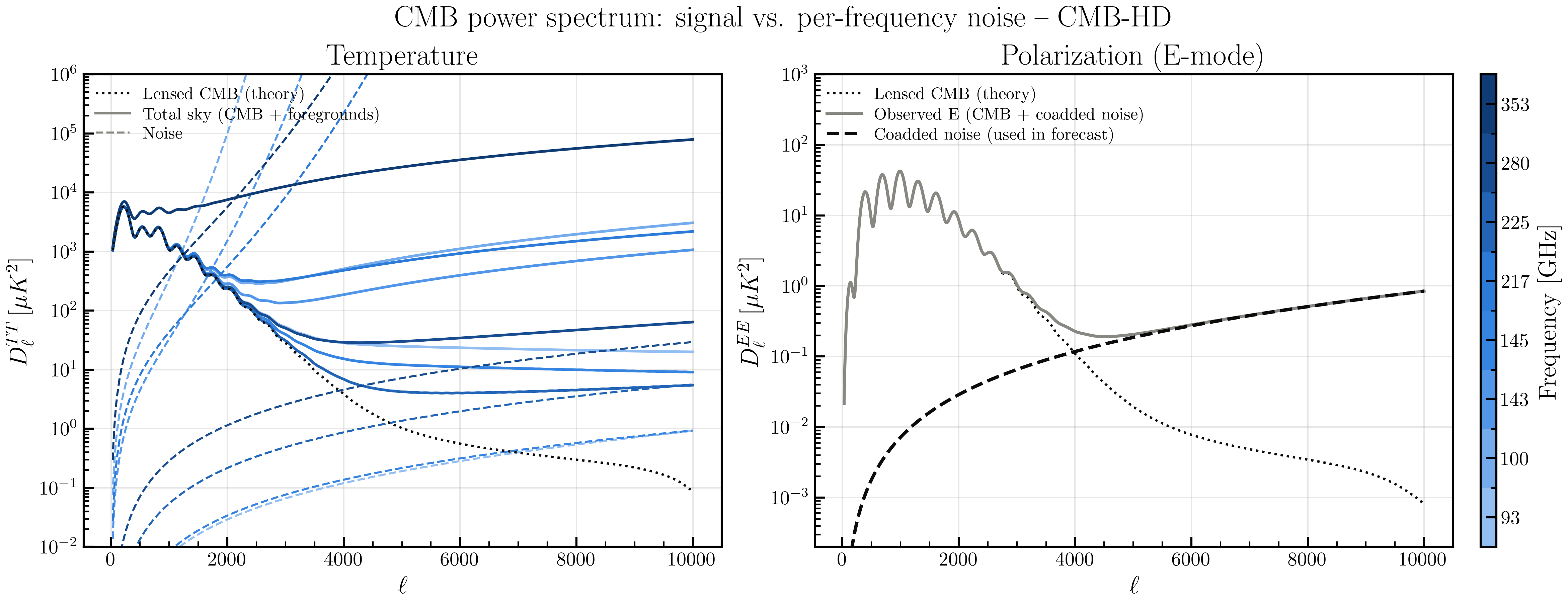}
    \caption{CMB temperature (left) and $E$-mode polarization (right) power spectra for the CMB-HD noise setup, for the same same frequency channels (93--353\,GHz, colorbar) and conventions as Figure~\ref{fig:cmb_spectra_SO}. The much lower CMB-HD instrument noise (dashed curves, left) extends the frequency-dependent foreground-dominated regime to substantially higher $\ell$ than for SO.}
    \label{fig:cmb_spectra_cmbhd}
\end{figure*}

\begin{figure}
    \centering
    \includegraphics[width=1.0\linewidth]{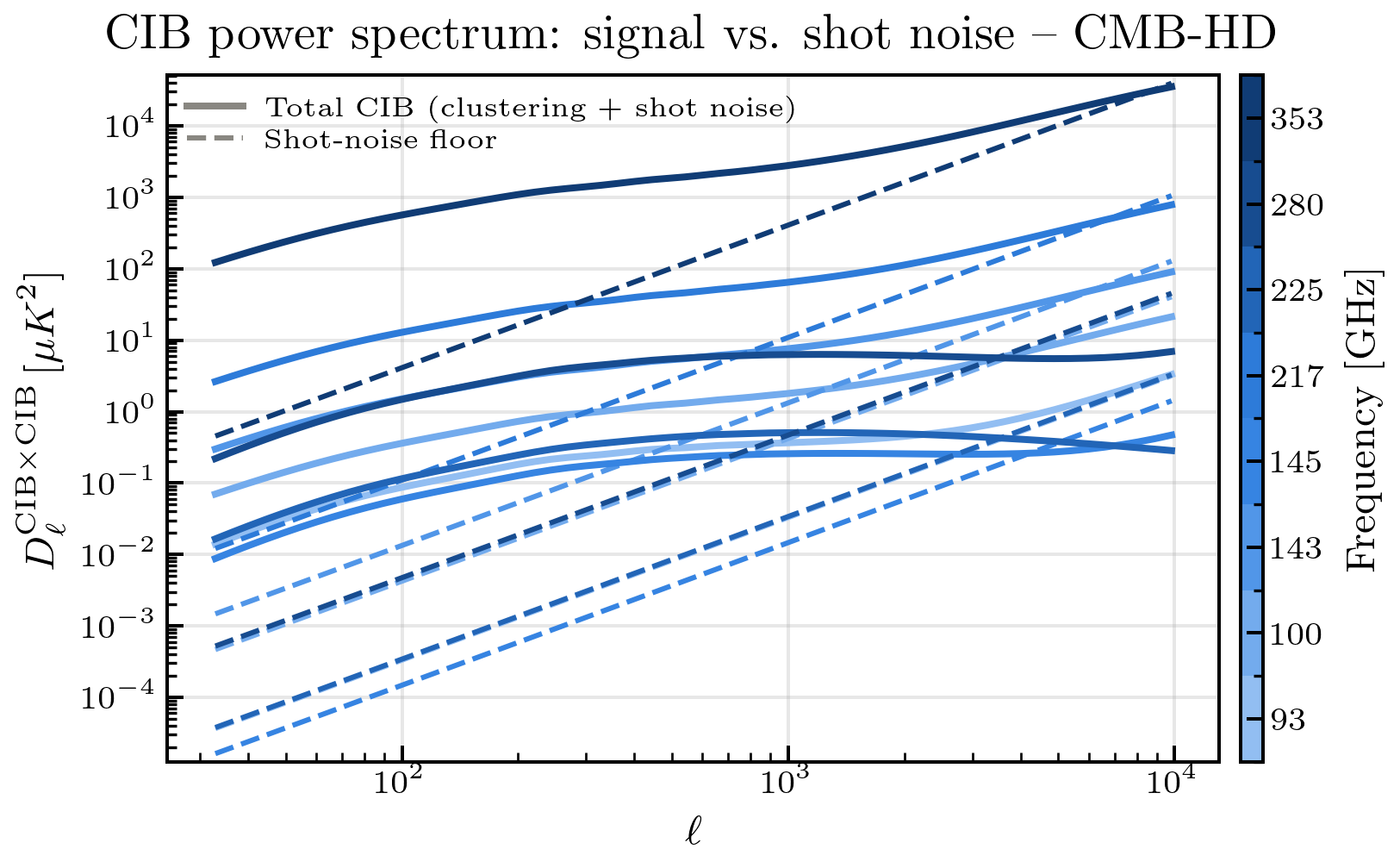}
    \caption{CIB auto-power spectra for the CMB-HD noise setup, for the same frequency channels and conventions as Figure~\ref{fig:cib_spectra_SO}. The Poisson shot-noise floor (dashed) is markedly lower than in the SO case at CMB-HD's native SO-band channels (93, 145, 225, 280\,GHz) --- by up to two orders of magnitude --- since CMB-HD's deeper point-source detection masks much fainter DSFGs; the four Planck-band channels (100, 143, 217, 353\,GHz) are unchanged between cases, as they use a fixed Planck-based flux cut independent of the noise model.}
    \label{fig:cib_spectra_cmbhd}
\end{figure}

Figures~\ref{fig:cmb_spectra_cmbhd} and \ref{fig:cib_spectra_cmbhd} show the CMB T/E and CIB power spectra for this CMB-HD setup. Relative to the SO case (Figures~\ref{fig:cmb_spectra_SO} and \ref{fig:cib_spectra_SO}), the lower instrument noise benefits both channels: in temperature, instrument noise no longer exceeds the residual foreground power at high $\ell$ as it does for SO, so a substantially larger range of scales becomes foreground-limited rather than noise-limited; in polarization, the noise floor likewise drops, although the underlying $E$-mode signal is unchanged since our pipeline includes no polarized-foreground model. For the range of multipoles, we limit ourselves to $\ell_{\rm min}=500$ for TT and ${\ell_{\rm min}}=300$ for both TE and EE, as the atmospheric complexity grows substantially at lower multipoles.  On the foreground side, the CIB (both clustering and shot noise) and radio Poisson power all change between cases, since CMB-HD's deeper point-source detection threshold masks much fainter sources, lowering the effective flux cut and thereby reducing the residual power in each component. The CIB shot-noise floor drops by up to two orders of magnitude at CMB-HD's native SO-band channels (93, 145, 225, and 280\,GHz). The four Planck-band channels (100, 143, 217, and 353\,GHz) are unaffected, retaining a fixed Planck-based flux cut independent of the noise model. CMB-HD therefore enters the joint TT+TE+EE Fisher forecast with both lower instrumental noise and a cleaner extragalactic foreground budget than SO, which we quantify in Section~\ref{sec:forecast_and_discussions}.

\section{Forecast and Discussion}
\label{sec:forecast_and_discussions}
We consider several cases for our forecasts, varying three ingredients independently. For the cosmological parameter set, we consider the base six $\Lambda$CDM parameters ($\omega_b$, $\omega_c$, $h$, $\tau$, $\ln(10^{10}A_s)$, $n_s$), as well as a seven-parameter extension that additionally lets the effective number of relativistic species $N_{\rm eff}$ float, since doing so shifts the parameter degeneracies and can change how much galaxy-tracer cleaning helps. We set the sum of neutrino mass $\sum_i m_{\nu_i}=0.06$ eV with a single eigenstate. We do not impose any additional Gaussian prior on the optical depth $\tau$ because we already include large-scale polarization information from Planck. For the CMB experiment, we adopt SO as our near-term baseline and CMB-HD as a futuristic case with substantially lower instrument noise (Section~\ref{sec:galaxy_catalogue}). For the galaxy survey, we use the unWISE catalog as our baseline tracer, together with an approximate rescaling of its shot noise to represent the deeper Rubin-like galaxy sample expected in the near future, and a hypothetical futuristic sample with $100\times$ lower shot noise (see Section~\ref{sec:futuristic_configuration}).

\subsection{Improvements on the ILC-cleaned T maps}

\begin{figure*}
    \centering
    \includegraphics[width=0.32\linewidth]{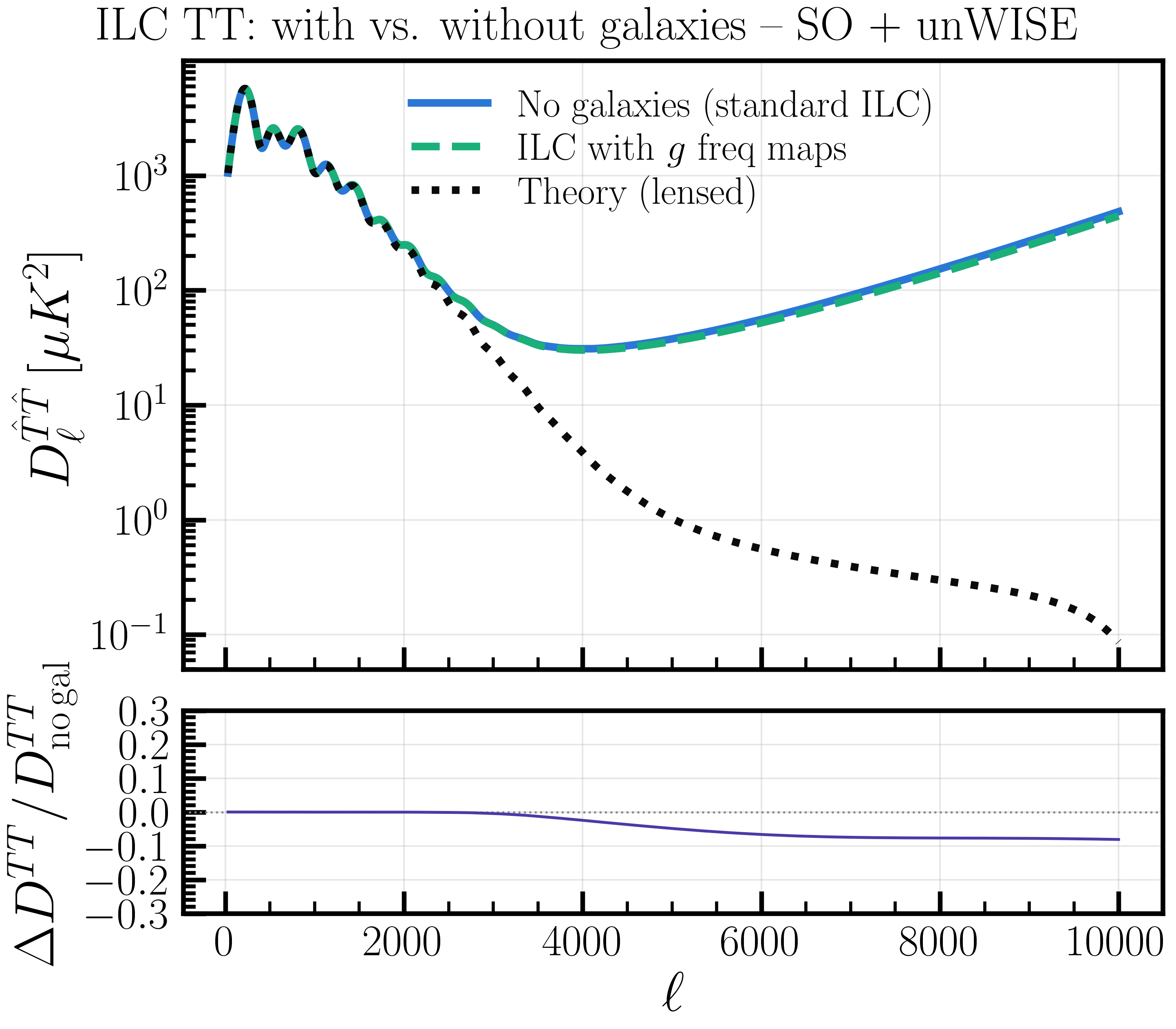}
    \includegraphics[width=0.32\linewidth]{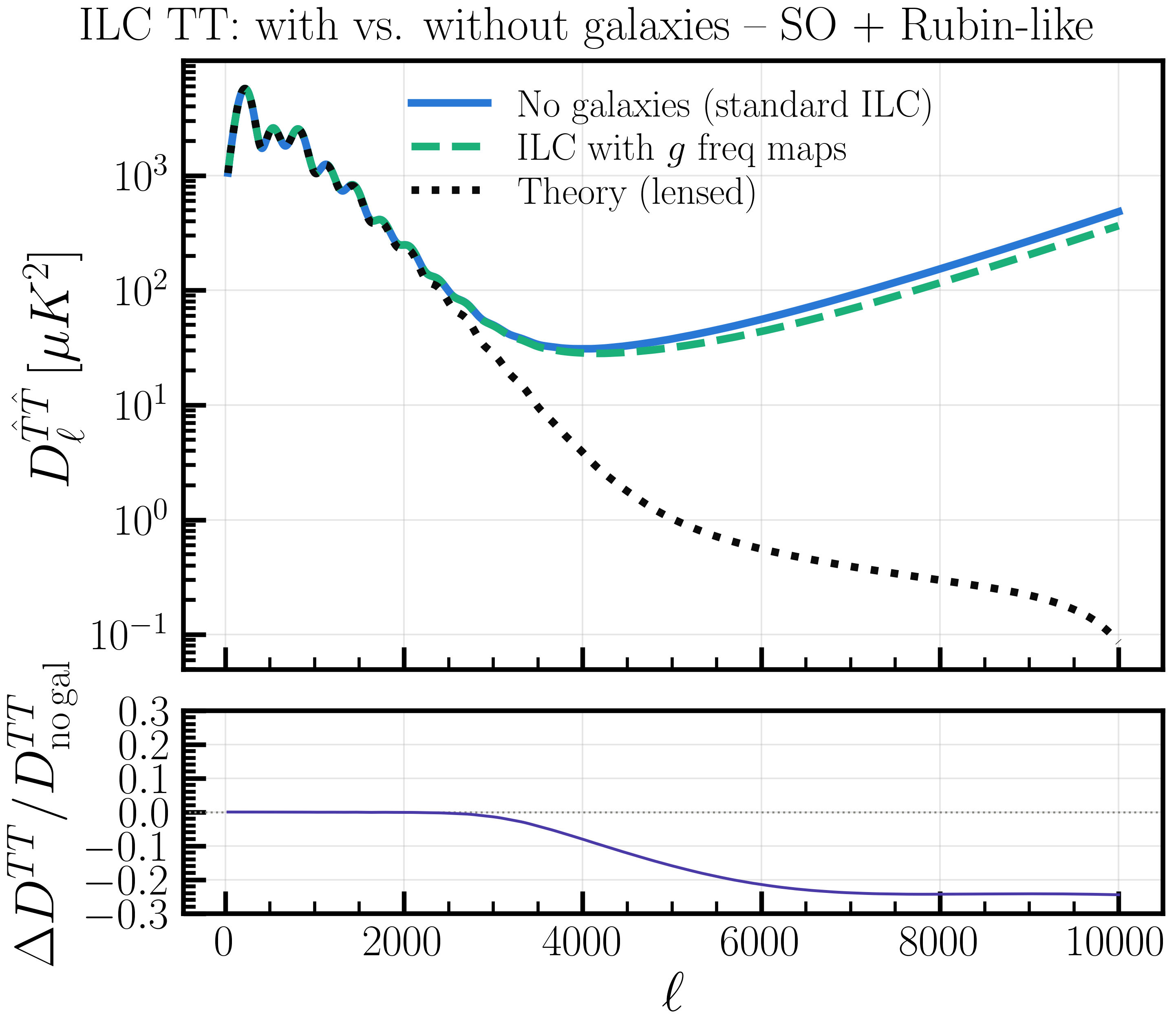}
    \includegraphics[width=0.32\linewidth]{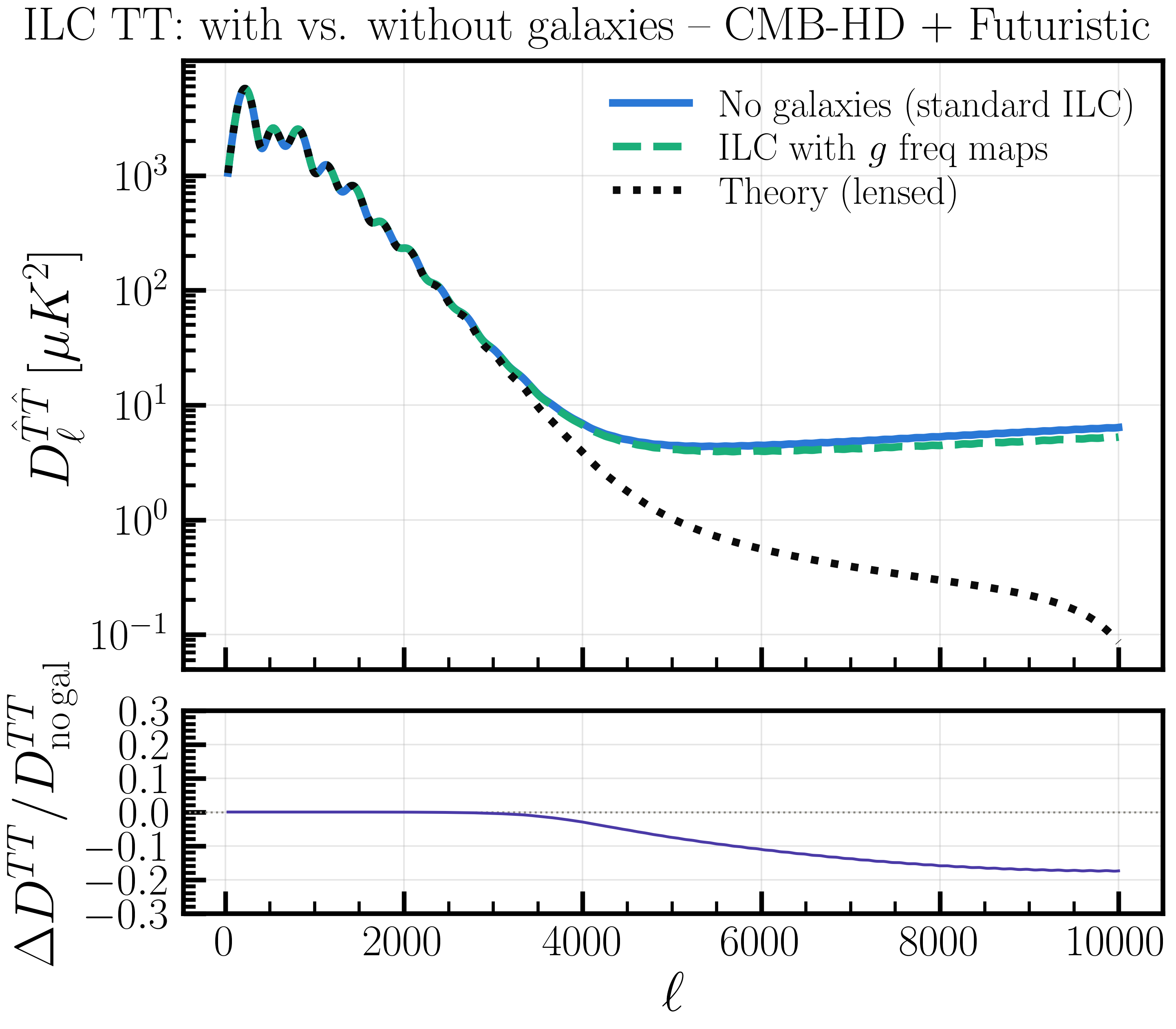}
    \caption{ILC-cleaned temperature power spectrum comparing three galaxy-sample/noise configurations: SO+unWISE (left), SO+Rubin-like (center), and CMB-HD+Futuristic (right). In each panel, the blue solid curve is the standard (temperature-only) ILC reconstruction; the green dashed curve adds galaxy-tracer maps to the ILC basis (g-freq ILC); the black dotted curve is the lensed-CMB theory spectrum. The lower sub-panel shows the fractional change from adding galaxies, $\Delta D_\ell^{TT}/D_\ell^{TT,\,\rm no\,gal}$, which reaches $\sim$--8\%, $\sim$--24\%, and $\sim$--17\% at $\ell\sim10^4$ for the unWISE, Rubin-like, and Futuristic samples respectively, reflecting the residual extragalactic foreground power removed by galaxy-tracer cleaning at high $\ell$ --- with the deeper Rubin-like and Futuristic tracers giving a substantially larger improvement than the current unWISE catalog.}
    \label{fig:ilc_comparison}
\end{figure*}

We first examine how adding galaxy tracers to the ILC changes the cleaned temperature power spectrum, before turning to their impact on cosmological parameter constraints in Section~\ref{sec:param_constraints}. Figure~\ref{fig:ilc_comparison} compares the standard, temperature-only ILC reconstruction to the galaxy-frequency (g-freq) ILC --- which additionally uses galaxy-tracer maps as cleaning templates --- across our three fiducial configurations.

In all three cases, the two ILC reconstructions are indistinguishable below $\ell\sim3000$, where the CMB itself dominates the total temperature power spectrum. The galaxy-cleaned spectrum only begins to depart from the standard ILC once extragalactic foregrounds become significant at smaller scales, and the size of this departure tracks the depth of the galaxy tracer rather than the sensitivity of the CMB experiment. The current unWISE catalog removes comparatively little residual foreground power ($\sim$4\% at $\ell\sim10^4$), while the denser Rubin-like and futuristic samples remove substantially more ($\sim$22\% and $\sim$32\%, respectively), consistent with a denser tracer more faithfully capturing the small-scale structure of the foregrounds it is meant to subtract. This same ordering does not carry over directly to the overall variance of the cleaned map, however: at $\ell\sim10^4$, adding galaxy tracers improves the total variance by 8\%, 24\%, and 17\% for the unWISE, Rubin-like, and futuristic combinations, respectively, so the Rubin-like case yields the largest variance improvement despite removing less raw foreground power than the futuristic sample. This reflects the fact that the variance improvement depends not only on how much foreground power the tracer removes, but also on how that residual compares to the instrument noise of the paired CMB experiment. We also note that differences in the flux cut for the two CMB experiments can also affect the fractional change in the overall variance.

\begin{figure*}
    \centering
    \includegraphics[width=0.32\linewidth]{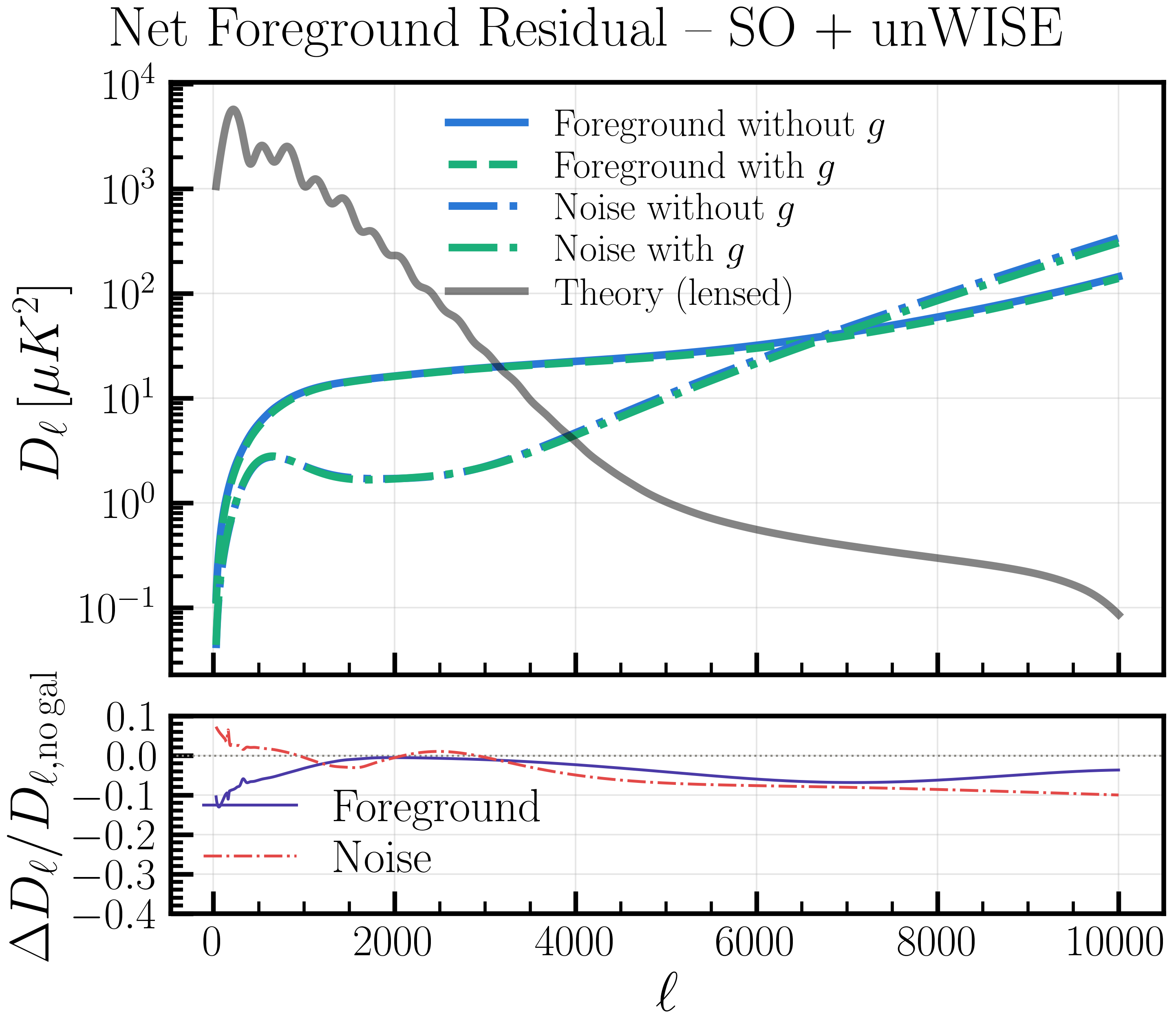}
    \includegraphics[width=0.32\linewidth]{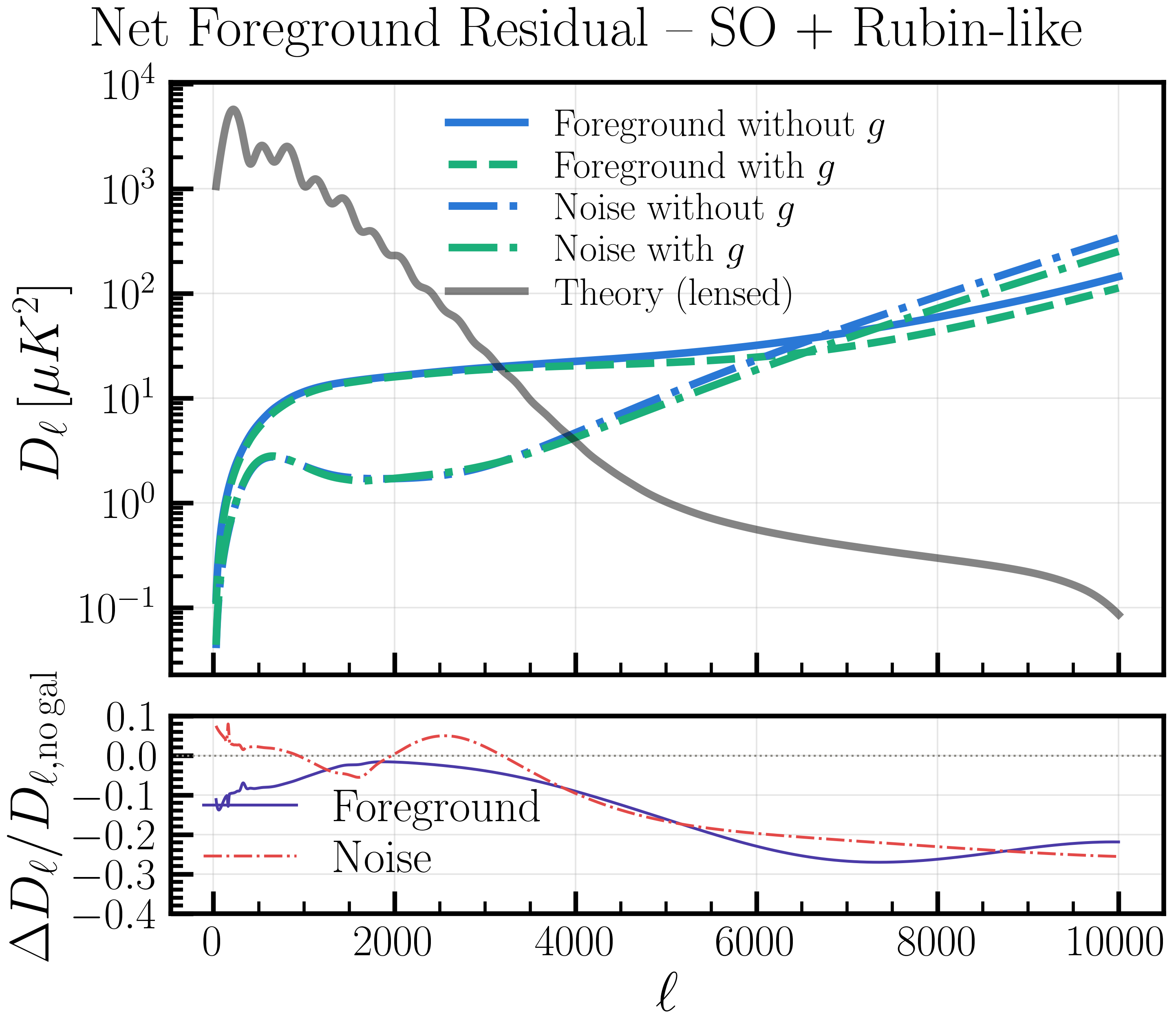}
    \includegraphics[width=0.32\linewidth]{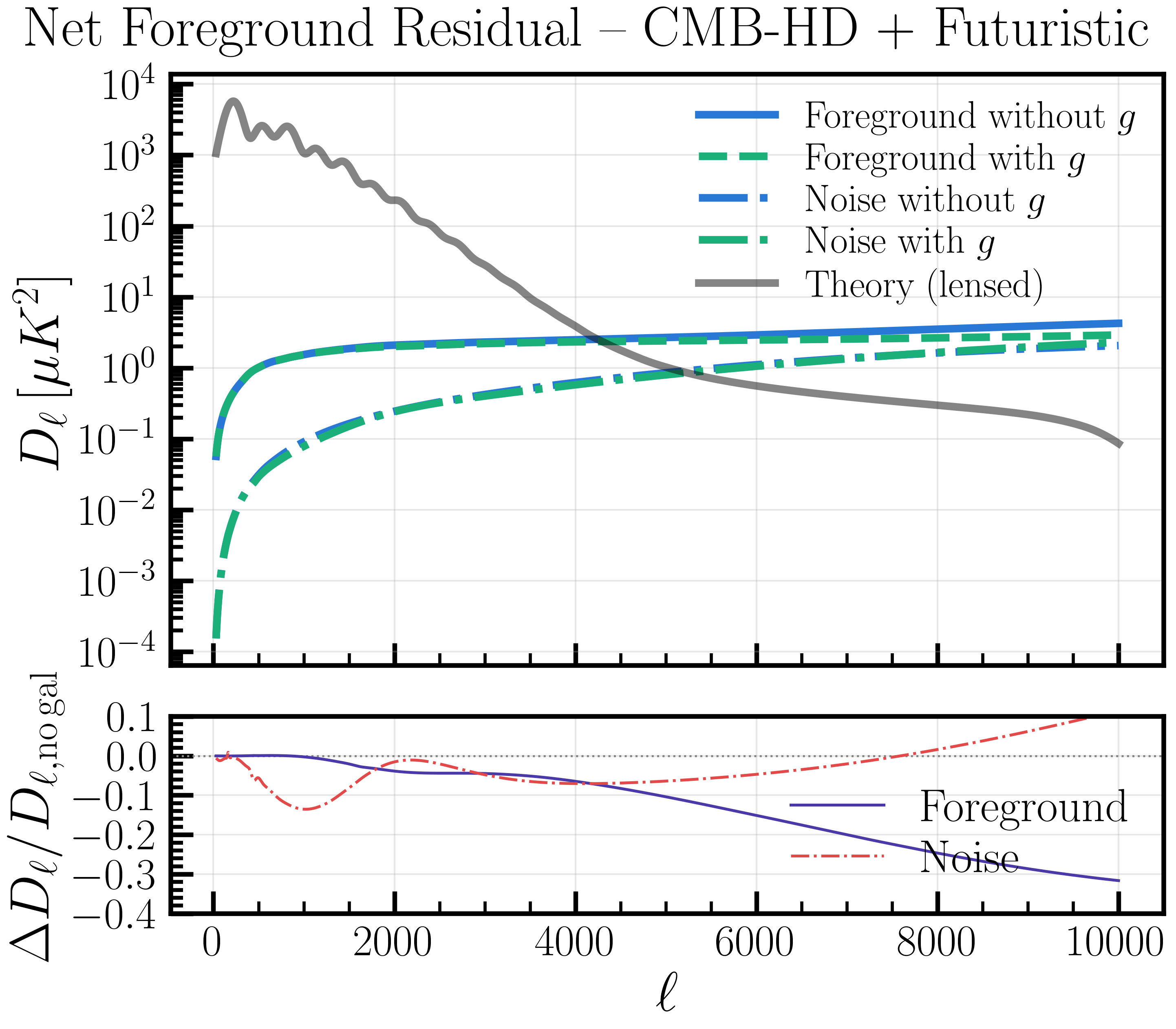}
    \caption{Residual foreground contamination and propagated instrument noise in the ILC-cleaned temperature map, with and without galaxy-tracer cleaning, comparing three galaxy-sample/noise configurations: SO+unWISE (left), SO+Rubin-like (center), and CMB-HD+Futuristic (right). In each panel, the blue solid curve is the residual foreground power for the standard (temperature-only) ILC reconstruction; the green dashed curve is the residual foreground power after adding galaxy-tracer maps to the ILC basis (g-freq ILC); the dash-dotted curve is the propagated instrument noise, computed from the actual ILC weights for each case; the lensed CMB theory spectrum is shown in gray for reference. The lower sub-panel shows the fractional change in the residual foreground power and the instrumental noise of CMB from adding galaxies, $\Delta D_\ell^{\rm fg}/D_\ell^{\rm fg,\,no\,gal}$, which reaches $\sim$--4\%, $\sim$--22\%, and $\sim$--32\% at $\ell\sim10^4$ for the unWISE, Rubin-like, and Futuristic samples respectively.} 
    \label{fig:residual_foreground}
\end{figure*}

Figure~\ref{fig:residual_foreground} shows the residual foreground power and the propagated instrument noise in the ILC-cleaned map, with and without the additional galaxy channels. For SO, the residual foreground dominates over the noise at intermediate scales but is overtaken by the noise at $\ell \gtrsim 7000$; for CMB-HD, by contrast, the foreground remains the dominant contaminant across the full range of scales considered. Notably, the galaxy tracers reduce the residual foreground power more effectively for CMB-HD than for SO+Rubin-like ($\sim$32\% and $\sim$22\% at $\ell \sim 10^4$, respectively), even though the corresponding gain in the total cleaned power spectrum is smaller for CMB-HD as shown in Figure~\ref{fig:ilc_comparison}.

\subsection{Effects on cosmological parameter constraints}
\label{sec:param_constraints}
A change in the cleaned power spectrum does not necessarily translate into a significantly tighter cosmological constraint. We therefore use a Fisher-matrix forecast to compare marginalized $1\sigma$ parameter uncertainties with and without galaxy tracers in the ILC basis, for both the base 6-parameter $\Lambda$CDM set and its $N_{\rm eff}$-extended 7-parameter counterpart, across all three configurations. In computing the Fisher matrix, the parameter derivatives are taken with respect to the pure lensed CMB power spectrum only, while the covariance is constructed from the ILC-cleaned power spectrum (see Eq.~\ref{eq.cov}). We compute this forecast for two cases: the cleaned intensity map alone (T only), and the joint T+E spectra (TT/TE/EE).

\begin{figure*}
    \centering
    \includegraphics[width=0.45\linewidth]{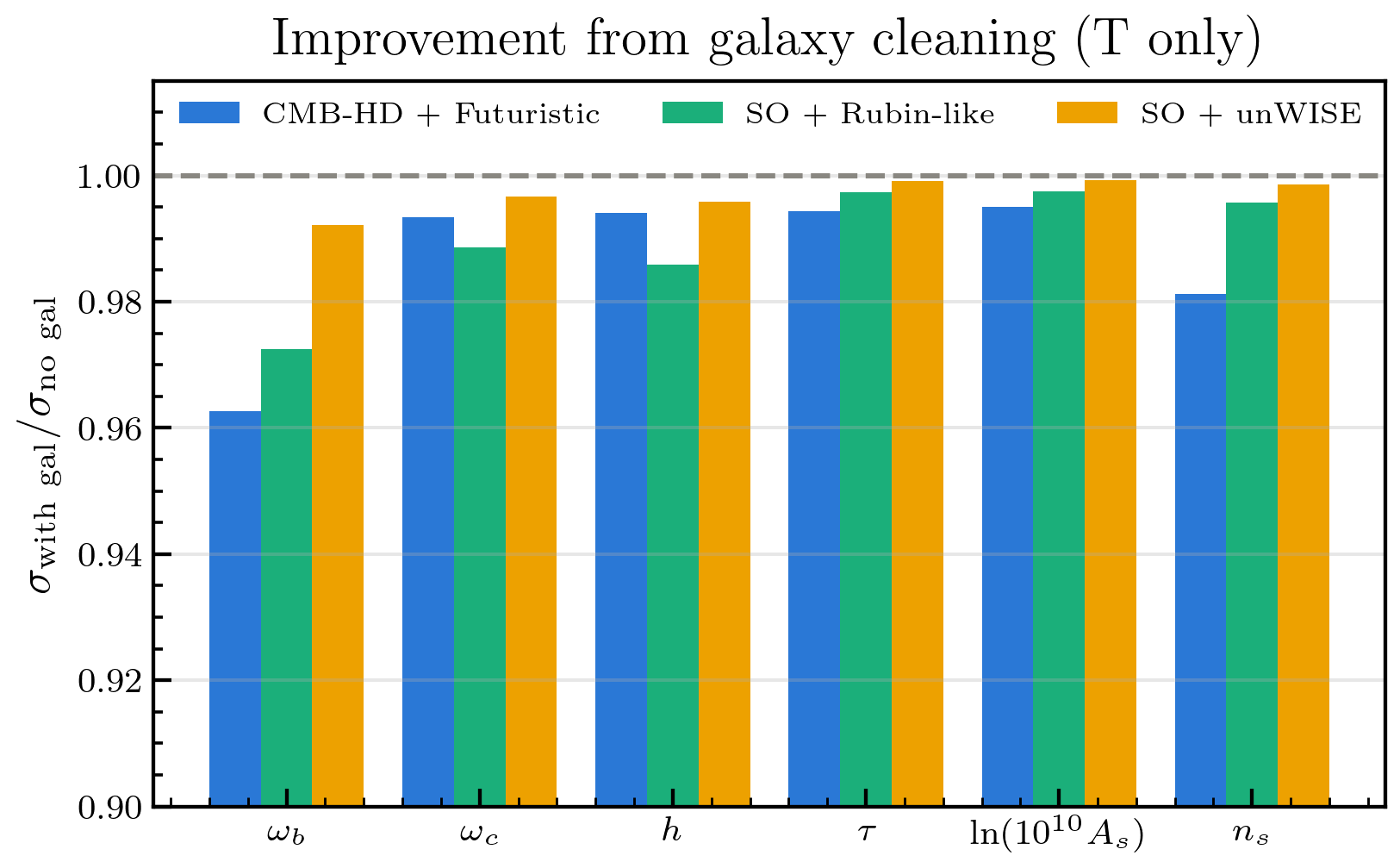}
    \includegraphics[width=0.45\linewidth]{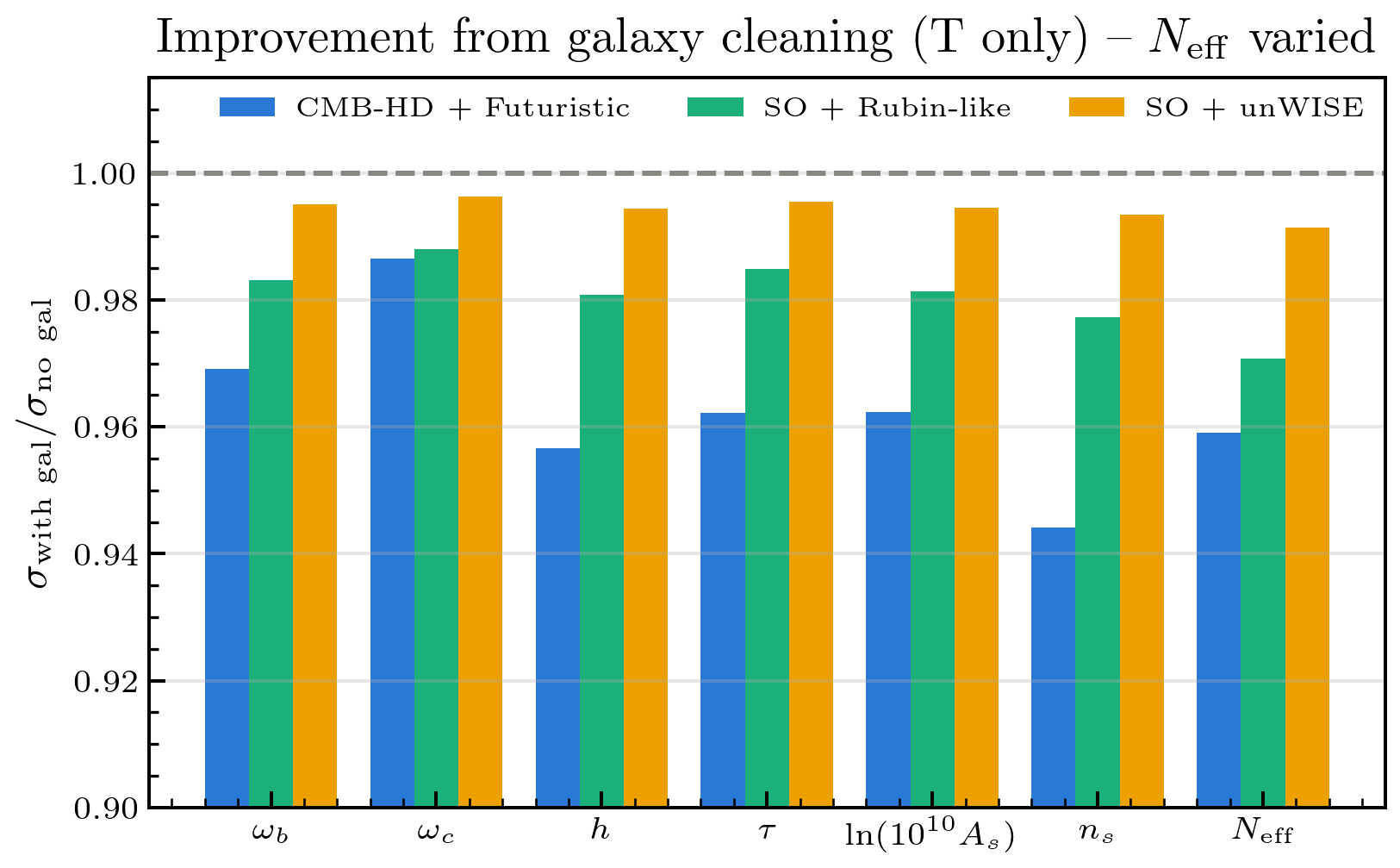}
    \caption{Ratio of marginalized 1$\sigma$ forecast uncertainties with vs.\ without galaxy tracers ($\sigma_{\rm with\,gal}/\sigma_{\rm no\,gal}$), from the T-only Fisher forecast over the full $\ell$ range, for the base 6-parameter $\Lambda$CDM set (left) and the 7-parameter set that additionally lets $N_{\rm eff}$ float (right). Each group of bars compares the three cases considered (CMB-HD+Futuristic, SO+Rubin-like, SO+unWISE); the dashed horizontal line marks unity, so bars below it indicate parameters for which adding galaxy tracers to the ILC tightens the constraint.}
    \label{fig:improvement_ratio_tt}
\end{figure*}

In the T-only forecast (Figure~\ref{fig:improvement_ratio_tt}), the galaxy tracer improves every base $\Lambda$CDM parameter across each case. In the nearly sample-variance-limited CMB-HD+Futuristic configuration, $\omega_b$ and $n_s$ tighten by $3.7\%$ and $1.9\%$, while the remaining parameters improve at the sub-percent level. In the more realistic SO+Rubin-like case, $\omega_b$, $\omega_{c}$, and $h$ tighten by $1.1$--$2.7\%$, while $\ln(10^{10}A_s)$ and $n_s$ remain at the sub-percent level. SO+unWISE follows the same trend, at roughly a quarter of the improvement seen for SO+Rubin-like.

\begin{figure*}
    \centering
    \includegraphics[width=0.45\linewidth]{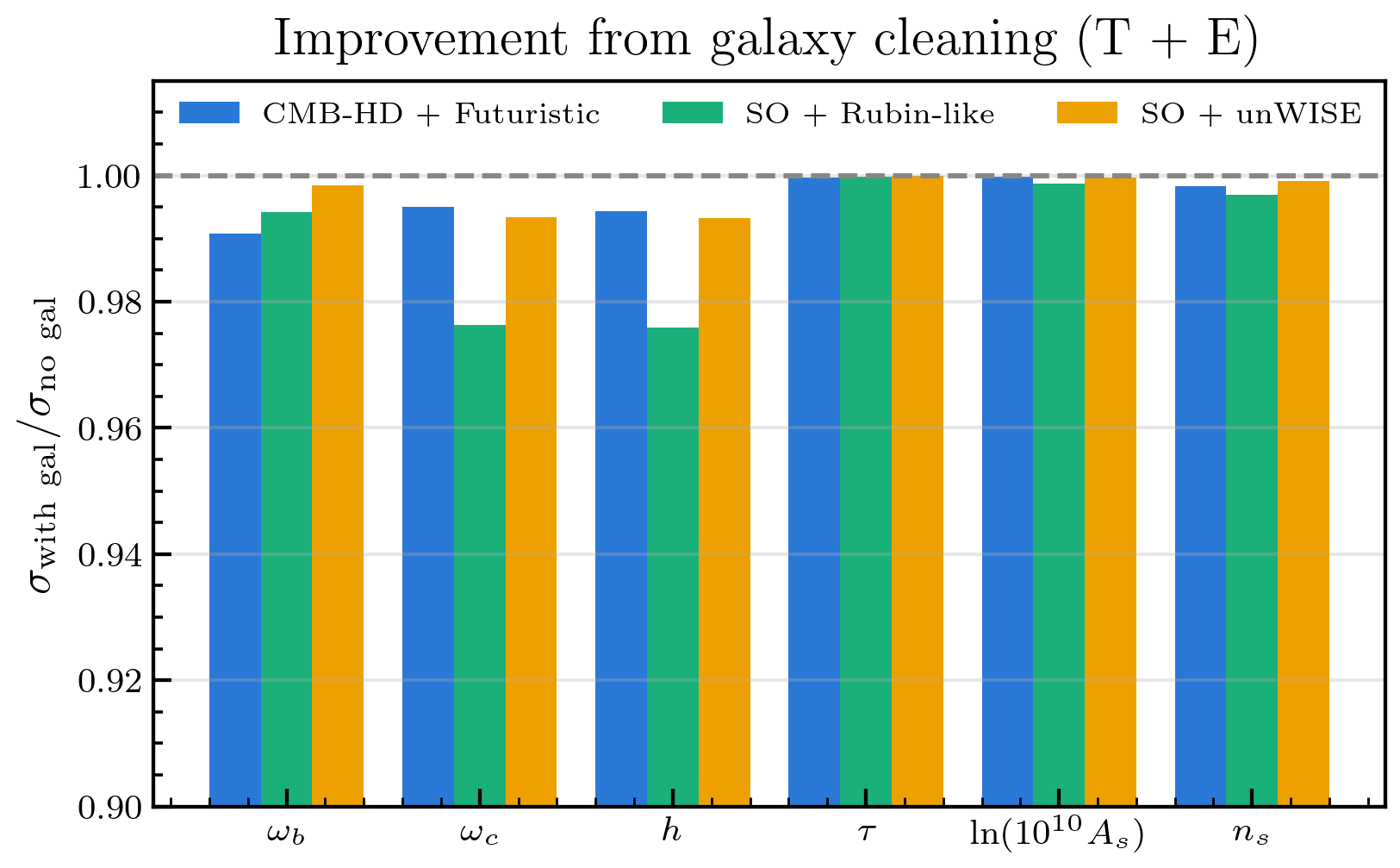}
    \includegraphics[width=0.45\linewidth]{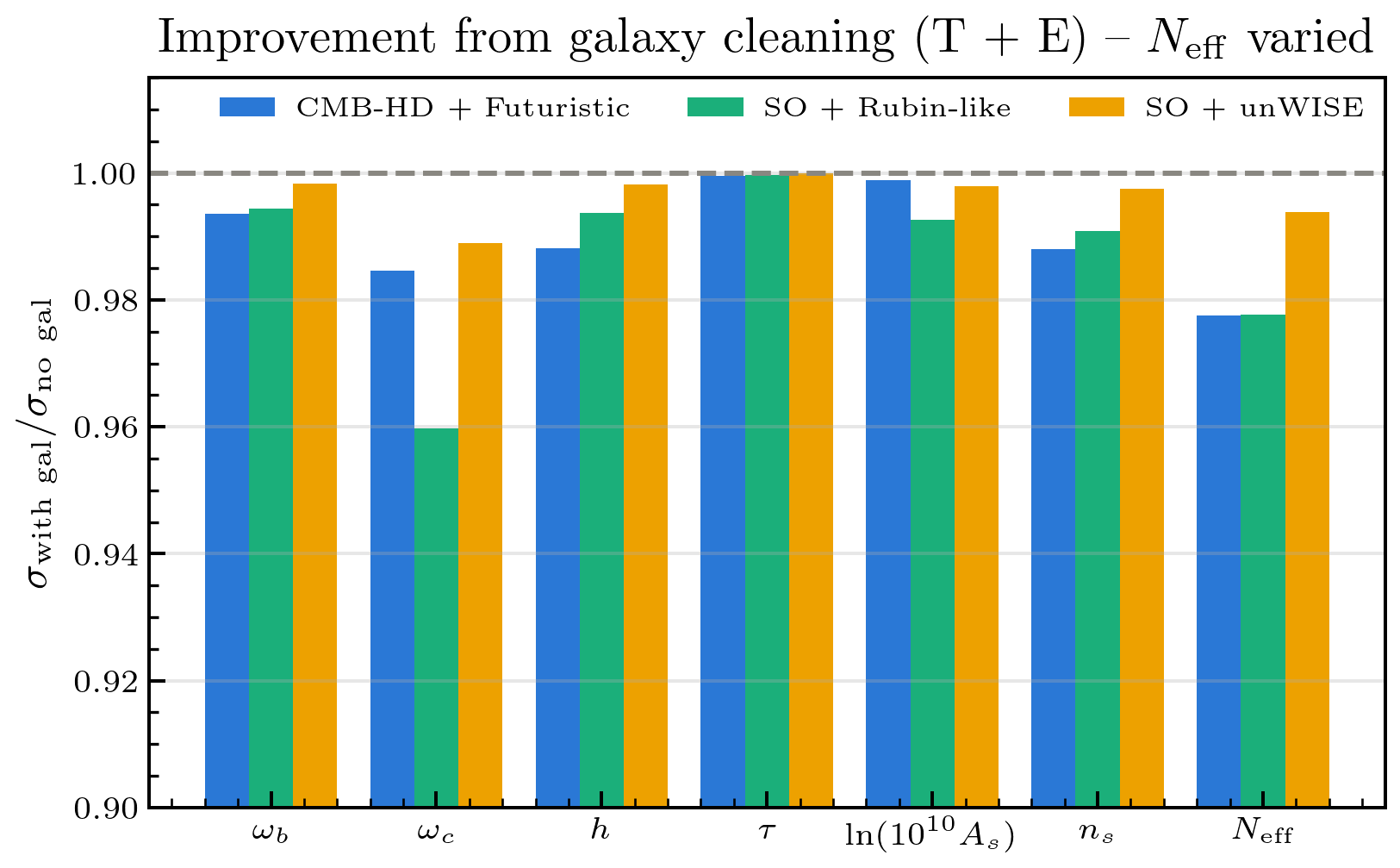}
    \caption{As in Figure~\ref{fig:improvement_ratio_tt}, but from the joint TT+TE+EE Fisher forecast: ratio of marginalized 1$\sigma$ forecast uncertainties with vs.\ without galaxy tracers ($\sigma_{\rm with\,gal}/\sigma_{\rm no\,gal}$), for the base 6-parameter $\Lambda$CDM set (left) and the 7-parameter set that additionally lets $N_{\rm eff}$ float (right). Each group of bars compares the three cases considered (CMB-HD+Futuristic, SO+Rubin-like, SO+unWISE); the dashed horizontal line marks unity, so bars below it indicate parameters for which adding galaxy tracers to the ILC tightens the constraint. The improvements are systematically smaller here than in the T-only forecast, since polarization data already constrain the base parameters independently of the T-leg foreground cleaning (although the cleaning reduces the TE error bars, in addition to TT).}
    \label{fig:improvement_ratio}
\end{figure*}

Despite the substantial differences in the cleaned power spectrum seen in Figure~\ref{fig:ilc_comparison}, the resulting improvement in the T+E parameter constraints is considerably more modest than in the T-only case above: at most a few percent, consistently smallest for the current unWISE catalog (sub-percent improvement for all six base parameters), and largest for SO+Rubin-like, where $\omega_{c}$ and $h$ tighten by $2.4\%$.  This likely arises from the substantial information on these parameters carried by the two-point lensing (peak-smearing), which is most prominent in TT. The same reduction holds for CMB-HD though there the improvements are all at the sub-percent level since its already small noise leaves little additional room for the galaxy tracer to improve and smaller fractional variance improvement on the power spectrum compared to SO+Rubin-like.

Allowing $N_{\rm eff}$ to vary, the improvement is similar between CMB-HD+Futuristic and SO+Rubin-like ($2.2\%$), and smallest for SO+unWISE ($0.6\%$) when considering T+E, tracking the LSS tracer depth and CMB instrument sensitivity directly. This is in contrast to the base $\Lambda$CDM parameters, whose fractional improvement is diluted once $N_{\rm eff}$ opens a near-degenerate direction, especially for the T-only case, which shows much wider constraints for the base $\Lambda$CDM parameters. We show the full cosmological constraints in Appendix~\ref{appendix:cosmological_constraints}.

\subsection{Prospects of subtracting ISW and kSZ}
We also study the prospects of subtracting the integrated Sachs-Wolfe (ISW) effect at large scales (see Refs.~\cite{Francis:2009pt,Mead:2010bv,Kim:2013nea,Muir:2016veb} for related studies). This subtraction can be performed via a simple linear combination,
\begin{align}
    \hat{T}_{\ell m}^\text{de-ISW} = \hat{T}_{\ell m} - \frac{C_{\ell}^{Tg}}{C_{\ell}^{gg}}\hat{g}_{\ell m}\,,
\end{align}
where $\hat{g}_{\ell m}$ is the projected galaxy overdensity field in harmonic space. The resulting power spectrum is
\begin{align}
    C_{\ell}^{T^\text{de-ISW}T^\text{de-ISW}} = C_\ell^{TT} - \left(\frac{C_\ell^{Tg}}{C_\ell^{gg}}\right)^2\,.
\end{align}
Even assuming a perfect ISW tracer, this method removes only $\sim35\%$ of the power at $\ell\sim\mathcal{O}(1)$, and this fraction decays quickly beyond $\ell=100$; as a result, we find only sub-percent improvements for all $\Lambda$CDM parameters.  Such methodology would be most useful in studying signals confined to very large scales in the CMB, such as primordial features.

A similar template-subtraction approach has been applied to the late-time kSZ signal in Ref.~\cite{Foreman:2022ves}, where separate Wiener-filtered estimates of the large-scale velocity and small-scale electron density fields are reconstructed from an external galaxy survey, combined, and projected along the line of sight to form a subtractable kSZ template. Using survey-specific halo occupation distributions, Ref.~\cite{Foreman:2022ves} forecast de-kSZing efficiencies of at most $\sim20\%$ for individual current and futuristic spectroscopic surveys, rising to $\sim35\%$ at $\ell\sim1000$ for an idealized combination of all of them. As with the ISW case above, the resulting gains in cosmological parameter constraints are nominally modest: even at a hypothetical $90\%$ de-kSZing efficiency, the uncertainties on $N_\text{eff}$ and $\Omega_b$ improve by no more than $\sim10\%$. Given how costly it is to shrink $\sigma(N_\text{eff})$ through survey depth, area, or resolution alone~\cite{CMB-S4:2016ple}, however, a $10$--$20\%$ reduction is still practically useful, and combined with the galaxy-tracer-assisted ILC gains reported above, points to a broader case for incorporating external large-scale-structure data into future CMB analyses beyond any single cleaning technique.

\section{Conclusions}
\label{sec:conclusions}
We have applied the galaxy-tracer-extended ILC framework of Ref.~\cite{Kusiak:2023hrz} to forecast the cosmological parameter gains achievable by incorporating LSS tracers into CMB temperature-map cleaning for near- and next-generation experiments. Using a halo-model foreground pipeline --- comprising CIB, tSZ, kSZ, and radio sources, calibrated against ACT and Planck measurements --- we computed ILC-cleaned temperature map power spectra and their cross-correlation with observed $E$-mode polarization, and used the resulting joint TT+TE+EE power spectra as the basis for a Fisher-matrix forecast across three experimental configurations: SO+unWISE, SO+Rubin-like, and a futuristic CMB-HD setup with a hypothetical $10\times$-deeper galaxy survey.

Our main findings are as follows: Adding galaxy tracers to the ILC reduces residual foreground power in the cleaned temperature map by $\sim$4\%, $\sim$22\%, and $\sim$32\% at $\ell\sim10^4$ for the unWISE, Rubin-like, and futuristic samples, respectively. The size of this reduction is driven by the number density of the galaxy tracer sample rather than the sensitivity of the CMB instrument, reflecting the fact that a denser tracer more faithfully captures the small-scale structure of the CIB and tSZ fields it is used to subtract. This foreground-power reduction does not translate directly into the overall variance of the cleaned map, however: at $\ell\sim10^4$, the total variance improves by $\sim$8\%, $\sim$24\%, and $\sim$17\% for the unWISE, Rubin-like, and futuristic combinations, respectively, so the Rubin-like case yields the largest variance improvement despite the futuristic sample removing more raw foreground power, since the variance improvement depends on how the residual foreground compares to the instrument noise, not on foreground removal alone. Despite these spectral-level improvements, the resulting gains in marginalized cosmological parameter constraints are modest for the base six-parameter $\Lambda$CDM model: sub-percent for SO+unWISE, and at most $\sim2\%$ for the combination of SO and Rubin-like tracer when polarization is included, since polarization already constrains the base parameters independently of T-leg foreground cleaning. This discrepancy between spectral- and parameter-level improvements arises because much of the foreground power reduced at high $\ell$ is already sub-dominant to cosmic variance and instrument noise at the level of parameter inference. The picture is also the same when $N_{\rm eff}$ is allowed to vary: the improvement is similar between CMB-HD+Futuristic and SO+Rubin-like ($\sim$2.2\%), and smallest for SO+unWISE ($\sim$0.6\%), tracking tracer depth and instrument sensitivity directly.

We also compared this galaxy-tracer ILC approach against two related template-subtraction techniques: direct subtraction of the integrated Sachs-Wolfe effect at low $\ell$, and de-kSZing of the late-time kinetic Sunyaev-Zel'dovich signal at high $\ell$ using external LSS surveys~\cite{Foreman:2022ves}. In both cases, even near-idealized template efficiencies translate into only sub-percent-to-few-percent gains in cosmological parameter constraints.

Taken together, these results suggest that the primary near-term benefit of galaxy-tracer-assisted ILC cleaning lies in producing cleaner maps rather than in directly tightening parameter constraints, with the latter becoming meaningful only when both the CMB instrument noise and the tracer shot noise are pushed well beyond current capabilities, or when extended parameter spaces with additional degeneracies are considered. Such cleaned maps and their cross-correlations with the polarization could prove useful for kSZ power spectrum measurements~\cite{Kusiak:2023hrz}, primordial feature constraints, improvements in CMB bispectrum estimation (for which T-leg foregrounds already pose a challenge~\cite{Hill:2018ypf,Coulton:2022wln}), and other applications.

Several directions remain for future work. The forecast presented here treats the foreground model as perfectly known; in practice, residual uncertainty in the CIB and tSZ power spectra will introduce additional systematics that could alter the effective gains. Extending the analysis to include realistic sky masks, map-level noise correlations, and a broader set of extended cosmologies --- including massive neutrinos, dynamical dark energy and primordial non-Gaussianity --- would provide a more complete picture of where this technique offers the greatest leverage. Cross-correlations beyond the TE spectrum, such as CMB lensing and galaxy-CMB lensing cross-spectra derived from the same cleaned maps, as well as higher N-point functions, represent another natural avenue for extracting additional cosmological information from the same data combination, although careful modeling of the impact of the foreground subtraction on the resulting signals would be necessary.

\section{Acknowledgments}
We thank Mathew Madhavacheril and Adri Duivenvoorden for useful discussions.  SFC and JCH acknowledge support from NASA grant 80NSSC24K1093 [ATP]. JCH also acknowledges support from NASA grant 80NSSC23K0463 [ADAP] and the Sloan Foundation.  We also acknowledge the computing resources from Columbia University’s Shared Research Computing Facility project, which is supported by NIH Research Facility Improvement Grant 1G20RR030893-01, and associated funds from the New York State Empire State Development, Division of Science Technology and Innovation (NYSTAR) Contract C090171, both awarded April 15, 2010.  This is not an official Simons Observatory collaboration paper.

\appendix
\section{Constraints On Cosmological Parameters} \label{appendix:cosmological_constraints}
Table~\ref{tab:sigma_base6_all} and Table~\ref{tab:sigma_neff_all} give the full marginalized 1$\sigma$ forecast uncertainties underlying the improvement ratios discussed in Section~\ref{sec:param_constraints}, for each of the three instrument/tracer combinations considered (SO+unWISE, SO+Rubin-like, CMB-HD+Futuristic). Table~\ref{tab:sigma_base6_all} shows constraints on the base 6-parameter $\Lambda$CDM model, while Table~\ref{tab:sigma_neff_all} shows the corresponding constraints for the 7-parameter extension in which $N_{\rm eff}$ is allowed to vary. Within each table, results are further broken out by row-block into the T-only and T+E Fisher forecasts, and into the case with the CMB lensing power spectrum $C_\ell^{\kappa\kappa}$ included. In each block, the ``-'' column gives the uncertainty from the CMB-only ILC, and the ``+unWISE'', ``+Rubin-like'', and ``+Futuristic'' columns give the uncertainty after including the corresponding galaxy tracer in the cleaning. Our forecasted constraints with lensing included agree well with the values reported in Table 2 of Ref.~\cite{SimonsObservatory:2025wwn} for parameters such as $N_\mathrm{eff}$ and $n_s$.

\begin{table*}
    \centering
    \begin{tabular}{lccccc}
        \hline
         & \multicolumn{3}{c}{SO} & \multicolumn{2}{c}{CMB-HD} \\
        Parameter & - & +unWISE & +Rubin-like & - & +Futuristic \\
        \hline
        \multicolumn{6}{l}{\textit{T-only}} \\
        \hline
        $10^{5}\,\omega_b$        & 10.8  & 10.8  & 10.5  & 7.98  & 7.68  \\
        $10^{4}\,\omega_c$        & 20.6  & 20.5  & 20.3  & 19.6  & 19.5  \\
        $10^{3}\,h$               & 8.17  & 8.13  & 8.05  & 7.52  & 7.47  \\
        $10^{3}\,\tau$            & 18.6  & 18.6  & 18.6  & 18.7  & 18.5  \\
        $10^{2}\,\ln(10^{10}A_s)$ & 3.24  & 3.24  & 3.23  & 3.30  & 3.28  \\
        $10^{3}\,n_s$             & 4.89  & 4.89  & 4.87  & 5.59  & 5.48  \\
        \hline
        \multicolumn{6}{l}{\textit{T+E}} \\
        \hline
        $10^{5}\,\omega_b$        & 3.95  & 3.94  & 3.93  & 2.19  & 2.17  \\
        $10^{4}\,\omega_c$        & 4.57  & 4.54  & 4.46  & 3.25  & 3.23  \\
        $10^{3}\,h$               & 1.79  & 1.78  & 1.75  & 1.22  & 1.21  \\
        $10^{3}\,\tau$            & 2.84  & 2.84  & 2.84  & 2.84  & 2.84  \\
        $10^{3}\,\ln(10^{10}A_s)$ & 5.01  & 5.01  & 5.00  & 4.88  & 4.88  \\
        $10^{3}\,n_s$             & 1.89  & 1.89  & 1.88  & 1.71  & 1.71  \\
        \hline
        \multicolumn{6}{l}{\textit{T+E, with CMB lensing}} \\
        \hline
        $10^{5}\,\omega_b$        & 3.87  & 3.87  & 3.86  & 2.16  & 2.14  \\
        $10^{4}\,\omega_c$        & 3.98  & 3.96  & 3.93  & 3.23  & 3.21  \\
        $10^{3}\,h$               & 1.56  & 1.56  & 1.55  & 1.21  & 1.21  \\
        $10^{3}\,\tau$            & 2.83  & 2.83  & 2.83  & 2.83  & 2.83  \\
        $10^{3}\,\ln(10^{10}A_s)$ & 4.91  & 4.91  & 4.91  & 4.84  & 4.84  \\
        $10^{3}\,n_s$             & 1.78  & 1.78  & 1.78  & 1.65  & 1.65  \\
        \hline
    \end{tabular}
    \caption{Marginalized $1\sigma$ forecast uncertainties with vs.\ without galaxy tracers, base 6-parameter $\Lambda$CDM, across temperature-only vs.\ T+E and no-lensing vs.\ with-lensing cases. The "-" column implies the constraints are from CMB-only.}
    \label{tab:sigma_base6_all}
\end{table*}
 
\begin{table*}
    \centering
    \begin{tabular}{lccccc}
        \hline
         & \multicolumn{3}{c}{SO} & \multicolumn{2}{c}{CMB-HD} \\
        Parameter & - & +unWISE & +Rubin-like & - & +Futuristic \\
        \hline
        \multicolumn{6}{l}{\textit{T-only}} \\
        \hline
        $10^{5}\,\omega_b$        & 43.9  & 43.6  & 43.1  & 42.5  & 41.2  \\
        $10^{4}\,\omega_c$        & 20.6  & 20.5  & 20.3  & 20.1  & 19.9  \\
        $10^{3}\,h$               & 32.6  & 32.4  & 32.0  & 27.9  & 26.7  \\
        $10^{3}\,\tau$            & 32.1  & 32.0  & 31.6  & 30.4  & 29.2  \\
        $10^{2}\,\ln(10^{10}A_s)$ & 6.26  & 6.23  & 6.14  & 5.98  & 5.76  \\
        $10^{3}\,n_s$             & 18.3  & 18.1  & 17.9  & 12.5  & 11.8  \\
        $10^{2}\,N_{\rm eff}$     & 24.0  & 23.8  & 23.3  & 18.2  & 17.5  \\
        \hline
        \multicolumn{6}{l}{\textit{T+E}} \\
        \hline
        $10^{5}\,\omega_b$        & 5.88  & 5.87  & 5.84  & 3.26  & 3.24  \\
        $10^{4}\,\omega_c$        & 8.41  & 8.32  & 8.07  & 3.75  & 3.69  \\
        $10^{3}\,h$               & 3.74  & 3.73  & 3.71  & 2.50  & 2.47  \\
        $10^{3}\,\tau$            & 2.86  & 2.86  & 2.86  & 2.87  & 2.86  \\
        $10^{3}\,\ln(10^{10}A_s)$ & 5.47  & 5.45  & 5.43  & 4.98  & 4.98  \\
        $10^{3}\,n_s$             & 3.27  & 3.27  & 3.24  & 2.51  & 2.48  \\
        $10^{2}\,N_{\rm eff}$     & 4.79  & 4.76  & 4.68  & 2.33  & 2.28  \\
        \hline
        \multicolumn{6}{l}{\textit{T+E, with CMB lensing}} \\
        \hline
        $10^{5}\,\omega_b$        & 5.84  & 5.83  & 5.81  & 3.25  & 3.23  \\
        $10^{4}\,\omega_c$        & 7.37  & 7.32  & 7.20  & 3.73  & 3.67  \\
        $10^{3}\,h$               & 3.66  & 3.66  & 3.64  & 2.42  & 2.40  \\
        $10^{3}\,\tau$            & 2.85  & 2.85  & 2.85  & 2.86  & 2.86  \\
        $10^{3}\,\ln(10^{10}A_s)$ & 5.31  & 5.31  & 5.30  & 4.97  & 4.96  \\
        $10^{3}\,n_s$             & 3.16  & 3.15  & 3.13  & 2.34  & 2.31  \\
        $10^{2}\,N_{\rm eff}$     & 4.54  & 4.52  & 4.47  & 2.25  & 2.20  \\
        \hline
    \end{tabular}
    \caption{Marginalized $1\sigma$ forecast uncertainties with vs.\ without galaxy tracers, 7-parameter case with $N_{\rm eff}$ varied, across temperature-only vs.\ T+E and no-lensing vs.\ with-lensing cases. The "-" column implies the constraints are from CMB-only.}
    \label{tab:sigma_neff_all}
\end{table*}




\bibliography{ref.bib}
\end{document}